\documentclass{aa} 
\usepackage[dvipsnames]{xcolor} 
\usepackage{graphicx, subfigure, amsmath, pifont}
\usepackage[breaklinks=true]{hyperref} 
\hypersetup{colorlinks=true,citecolor=Blue}
\usepackage{natbib}
\bibpunct{(}{)}{;}{a}{}{,} 
\usepackage[english]{babel}
\usepackage{blindtext}
\usepackage{pifont}
\usepackage[normalem]{ulem}

\begin{document} 

 \title{A bright inner disk and structures in the transition disk around the very low-mass star CIDA\,1}

   \author{
   P.~Pinilla\inst{1,2}, N.~T.~Kurtovic \inst{1}, M.~Benisty\inst{3, 4}, C.~F.~Manara\inst{5}, A.~Natta\inst{6}, E.~Sanchis\inst{5,7}, M.~Tazzari\inst{8}, S.~M.~Stammler\inst{7}, L.~Ricci\inst{9}, and L.~Testi\inst{5, 10}
   }
   \institute{Max-Planck-Institut f\"{u}r Astronomie, K\"{o}nigstuhl 17, 69117, Heidelberg, Germany, \email{pinilla@mpia.de}
   \and Mullard Space Science Laboratory, University College London, Holmbury St Mary, Dorking, Surrey RH5 6NT, UK.
   \and Unidad Mixta Internacional Franco-Chilena de Astronom\'{i}a, CNRS/INSU UMI 3386, Departamento de Astronom\'ia, Universidad de Chile, Camino El Observatorio 1515, Las Condes, Santiago, Chile.
   \and Univ. Grenoble Alpes, CNRS, IPAG, 38000 Grenoble, France.
   \and European Southern Observatory, Karl-Schwarzschild-Str. 2, 85748 Garching, Germany.
   \and Dublin Institute for Advanced Studies, School of Cosmic Physics, 31 Fitzwilliam Place, Dublin 2, Ireland.
   \and University Observatory, Faculty of Physics, Ludwig-Maximilians-Universität München, Scheinerstr. 1, 81679 Munich, Germany.
   \and Institute of Astronomy, University of Cambridge, Madingley Road, CB3 0HA Cambridge, UK.
   \and Department of Physics and Astronomy, California State University Northridge, 18111 Nordhoff Street, Northridge, CA 91130, USA.
   \and INAF-Arcetri, Largo E. Fermi 5, 50125 Firenze, Italy.}
   \date{}

  \abstract
  {The frequency of Earth-sized planets in habitable zones appears to be higher around M-dwarfs, making these systems exciting laboratories to investigate planet formation. Observations of protoplanetary disks around very low-mass stars and brown dwarfs remain challenging and little is known about their properties. The disk around CIDA\,1 ($\sim$0.1-0.2\,$M_\odot$) is one of the very few known disks that host a large cavity (20\,au radius in size) around a very low-mass star. We present new ALMA observations at Band\,7 (0.9\,mm) and Band\,4 (2.1\,mm) of CIDA\,1 with a resolution of $\sim 0.05''\times 0.034''$. These new ALMA observations reveal a very bright and unresolved inner disk, a shallow spectral index of the dust emission ($\sim2$), and a complex morphology of a ring located at 20\,au. We also present X-Shooter (VLT) observations that confirm the high accretion rate of CIDA\,1 of $\dot{M}_{\rm acc}$ = 1.4 $~\times~10^{-8} M_\odot$/yr. This high value of $\dot{M}_{\rm acc}$, the observed inner disk, and the large cavity of 20\,au exclude models of photo-evaporation to explain the observed cavity.  When comparing these observations with models that combine planet--disk interaction, dust evolution, and radiative transfer, we exclude planets more massive than 0.5\,$M_{\rm{Jup}}$ as the potential origin of the large cavity because with these it is difficult to maintain a long-lived and bright inner disk. Even in this planet mass regime, an additional physical process may be needed to stop the particles from migrating inwards and to maintain a bright inner disk on timescales of millions of years. Such mechanisms include a trap formed by a very close-in extra planet or the inner edge of a dead zone. The low spectral index of the disk around CIDA\,1 is difficult to explain and challenges our current dust evolution models, in particular processes like fragmentation, growth, and diffusion of particles inside pressure bumps.}


   \titlerunning{ALMA Observations of CIDA\,1}
   \authorrunning{P.~Pinilla et al.}
   \maketitle

%
\section{Introduction}                  \label{sect:intro}
The current exoplanet population shows a large diversity of properties, which may have an equivalent in the properties and initial conditions of the parental protoplanetary disks. In the context of exoplanets, some trends have already emerged with stellar mass, such as giant planets being more frequent around more massive and more metal-rich stars, but sub-Neptunes being more frequent around lower mass stars \citep[e.g.,][]{santos2000, johnson2010, mulders2015}. In the context of protoplanetary disks, surveys  at millimeter wavelengths have shown that dust disk masses also increase with stellar mass \citep[e.g.,][]{andrews2013, pascucci2016, ansdell2017}. These two observational results are consistent with the predictions of core-accretion models for the formation of planets, in which giant planet formation is relatively inefficient in disks with low mass around very low-mass stars \citep[e.g.,][]{payne2007, liu2020}. Alternatively, giant planets around M-dwarfs may form when the disk is gravitationally unstable \citep[][]{mercer2020}. 

Recent observations of protoplanetary disks at high angular resolution and sensitivity, in particular with the Atacama Large Millimeter/submillimeter
Array (ALMA), revealed that nearly all large disks ($\gtrsim$50\,au in dust continuum emission) host substructures \citep[][]{andrews2018, long2018}, with rings and gaps being the most common type.  If these substructures are created by planets, most of these planets have masses and orbital radii that are not detectable by any of the methods currently used to detect exoplanets \citep[][]{bae2018, lodato2019}.

However, most of the current ALMA programs attempt to study the prevalence and diversity of these structures around T Tauri and Herbig AeBe stars with different properties; little is known about the existence of substructures in disks around very low-mass stars (VLMSs, $\lesssim$0.1-0.2\,$M_\odot$) or brown dwarfs (BDs) because observations of these disks are challenging as they are fainter and smaller \citep[e.g.,][]{rilinger2019, sanchis2020}. Recently, \cite{kurtovic2021} presented a sample of six disks around VLMSs in the Taurus star forming region observed at high resolution with ALMA, demonstrating that these disks follow the expected relations for disk dust masses versus stellar mass ($M_{\rm{dust}}\propto M_\star^{1.1}$) and disk dust radii versus disk luminosity ($R_{\rm{dust}}\propto L_{\rm{mm}}^{0.54}$) as seen in their counterparts \citep[more massive and larger disks, e.g.,][]{andrews2013, pascucci2016, hendler2020}. These recent observations are also unveiling that substructures in the low-mass star regime could be as common as in T-Tauri and Herbig disks (50\% of the sample revealed substructures at a resolution of 0.1'', although the sample was selected to target the brightest disks around VLMSs). These observations suggest that structures may be universal and they question the origin of the observed structures; for example, whether or not these rings and gaps may be formed by (giant) planets independent of stellar mass.

The spectral index, which is the slope of the spectral energy distribution (SED) at the millimeter wavelength, can be interpreted in terms of grains size with low values ($\lesssim 3.0$) corresponding to large grains \citep[e.g.,][]{draine2006}. Disks around VLMSs and BDs are known to host millimeter (mm)-sized particles, as evidenced by their low millimeter spectral indices \citep[e.g.,][]{ricci2014, pinilla2017}. These observations challenge current models of grain growth in protoplanetary disks, where radial drift velocities of dust particles towards the star are higher in VLMSs and BDs than around higher mass young stars \citep[][]{pinilla2013, zhu2018}. To retain enough particles in VLMS disks and explain low spectral index values, multiple pressure bumps of high amplitude are required. This would translate into multiple rings in millimeter observations with infinite angular resolution. Vortices can also trap particles \citep[e.g.,][]{barge1995, klahr2006, varniere2006, ataiee2013}, in which case observations would reveal one or more nonaxisymmetric features. Nonetheless, from the observational point of view, our current understanding of substructures around VLMSs is very limited, with only two clear cavities detected in CIDA\,1 \citep{pinilla2018} and MHO\,6 \citep{kurtovic2021} and a hint of rings and gaps in the disk around 2MASS04334465$+$2615005 \citep{kurtovic2021}.

CIDA\,1 (2MASS J04141760+2806096) is a young VLMS in the Taurus star forming region that is close to the hydrogen burning limit. It is known to be surrounded by a very luminous (for its mass) disk at millimeter emission. For this reason it was selected as a candidate to measure the spectral index and constrain the presence of large dust grains. ALMA Cycle 0 observations at 0.89 and 3.4mm revealed a shallow integrated spectral index ($\sim2$) and relatively large disk \citep{ricci2014}, suggesting the presence in the disk of a substantial population of millimeter-sized grains. ALMA Cycle 3 observations of CIDA\,1 revealed for the first time a dust-depleted inner cavity around a VLMS, with a ring with maximum brightness at $\sim$20 au \citep{pinilla2018}, which is a typical feature of the transition disk (TD) population. The observed structure in CIDA\,1 may suggest that cavity opening processes are common in millimeter-bright disks independent of the central star mass, including VLMSs \citep{pinilla2018b, pinilla2020, sinclair2020}.

The  mechanisms that are most likely to form the cavity in the CIDA\,1 disk include the interaction of the disk with a massive embedded planet. However, core-accretion models  predict that it is very challenging to create the required pressure bump with an embedded planet under the physical conditions of CIDA\,1, because a large fraction of the disk mass (5\%-45\%, where higher values correspond to a disk with high viscosity) is needed to form such a massive planet in the first place. One alternative is that the initial disk mass is high enough such that the disk is gravitationally unstable and capable of forming massive planets \citep[][]{mercer2020}. In the case of CIDA\,1, a Saturn-mass planet is the minimum mass required to open a gap (with low viscosity) and to trap millimeter-sized particles, but the predicted integrated spectral indices for this scenario are too high compared to observations \citep{pinilla2017}. This contrasts with T-Tauri and Herbig disks, where a single massive planet is capable of opening the cavity, and trapping and retaining enough mm-grains to explain low spectral indices \citep{pinilla2014}. As a consequence, it is expected that strong and multiple pressure bumps are present in VLMS disks with low spectral indices.

\begin{figure*}
 \centering
 \tabcolsep=0.05cm 
   \begin{tabular}{cc}   
        \includegraphics[width=9.0cm]{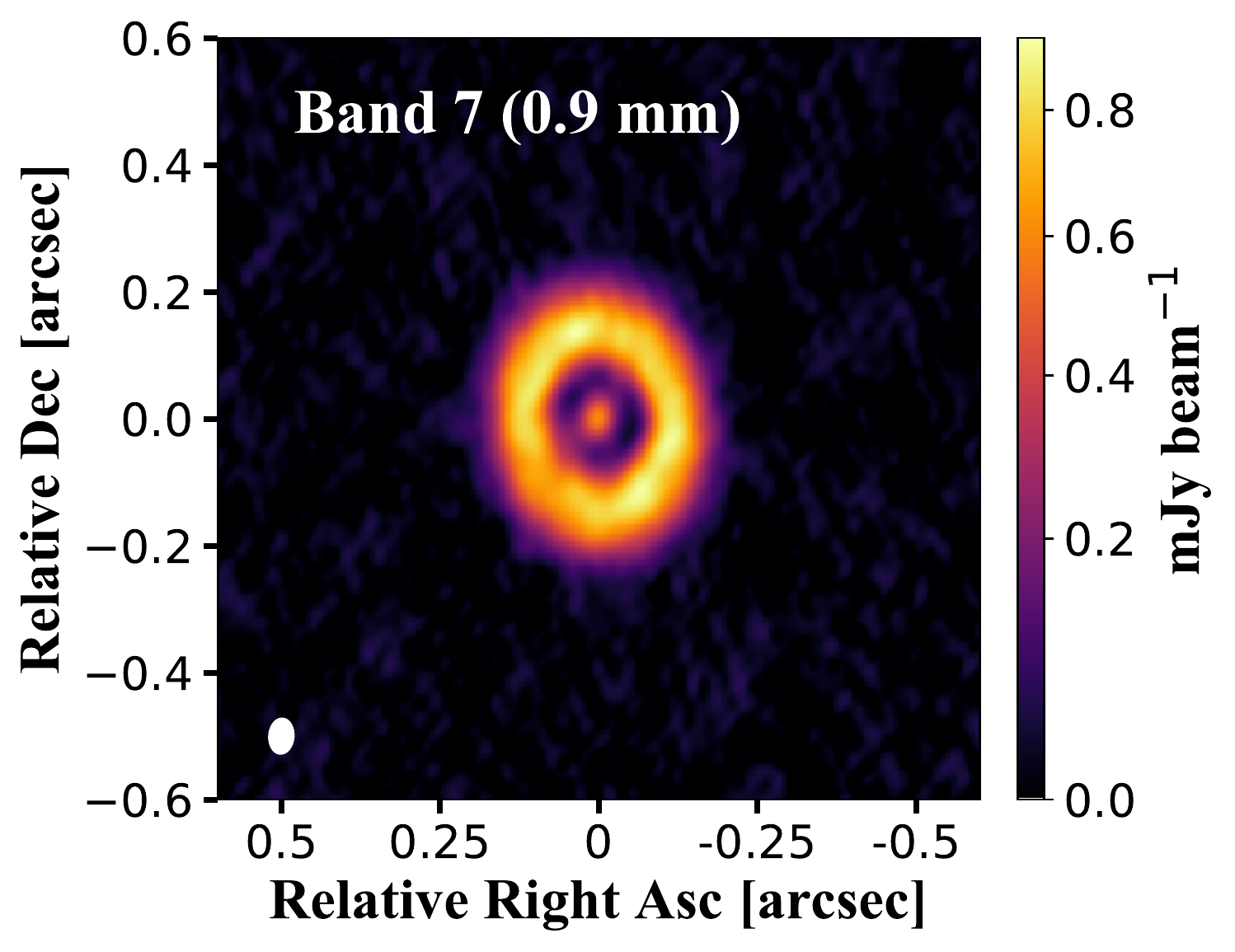}&
        \includegraphics[width=9.0cm]{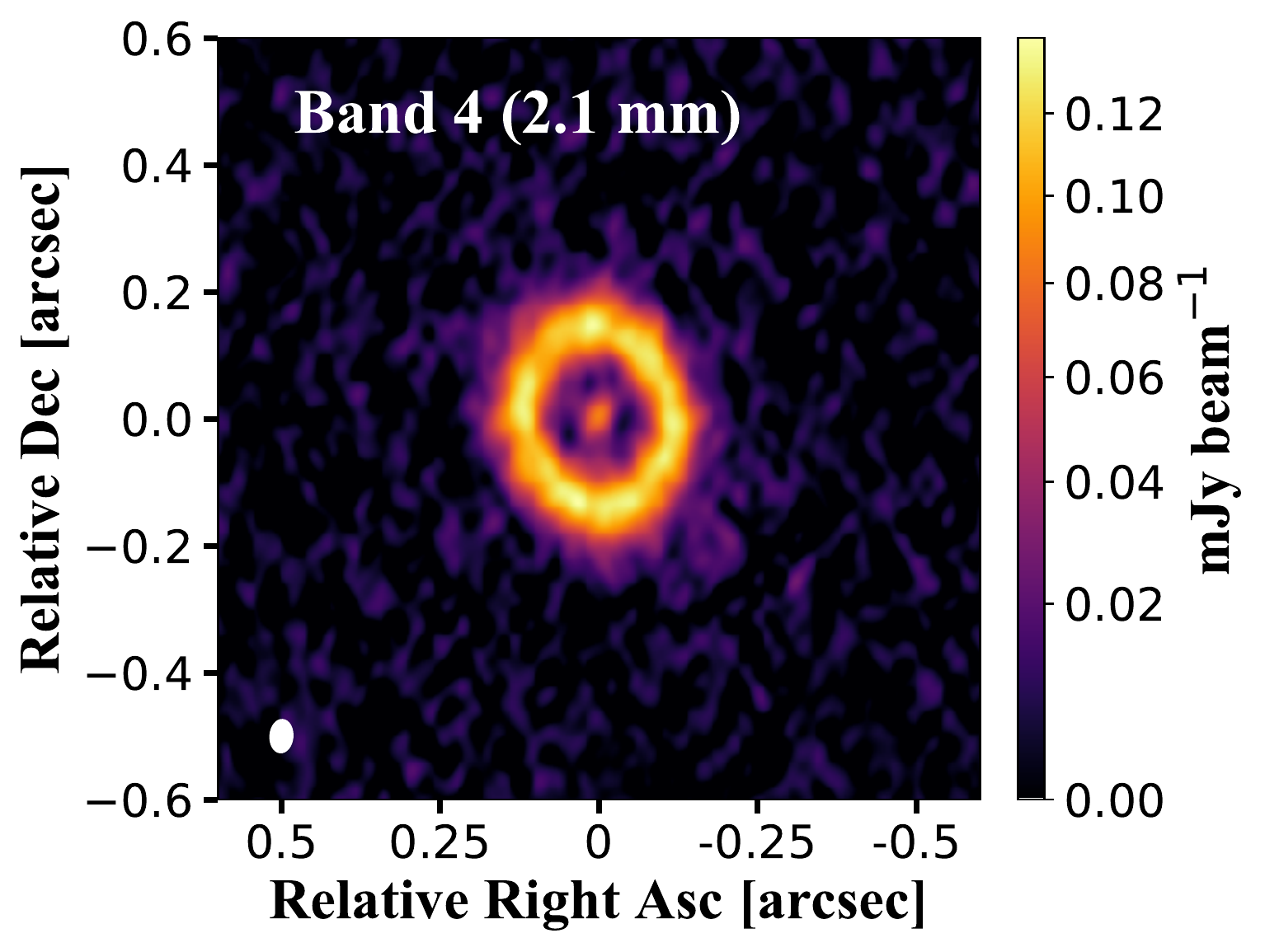}
    \end{tabular}
   \caption{ALMA Band 7 (0.9\,mm) and Band 4 (2.1 mm) continuum images of the CIDA\,1 disk. The synthesized beam is shown in the lower left corner of each image. The details of the images are summarized in Table~\ref{tab:summary_ALMA}.}
   \label{fig:ALMA_CIDA}
\end{figure*}

\begin{figure*}
   \includegraphics[width=18.0cm]{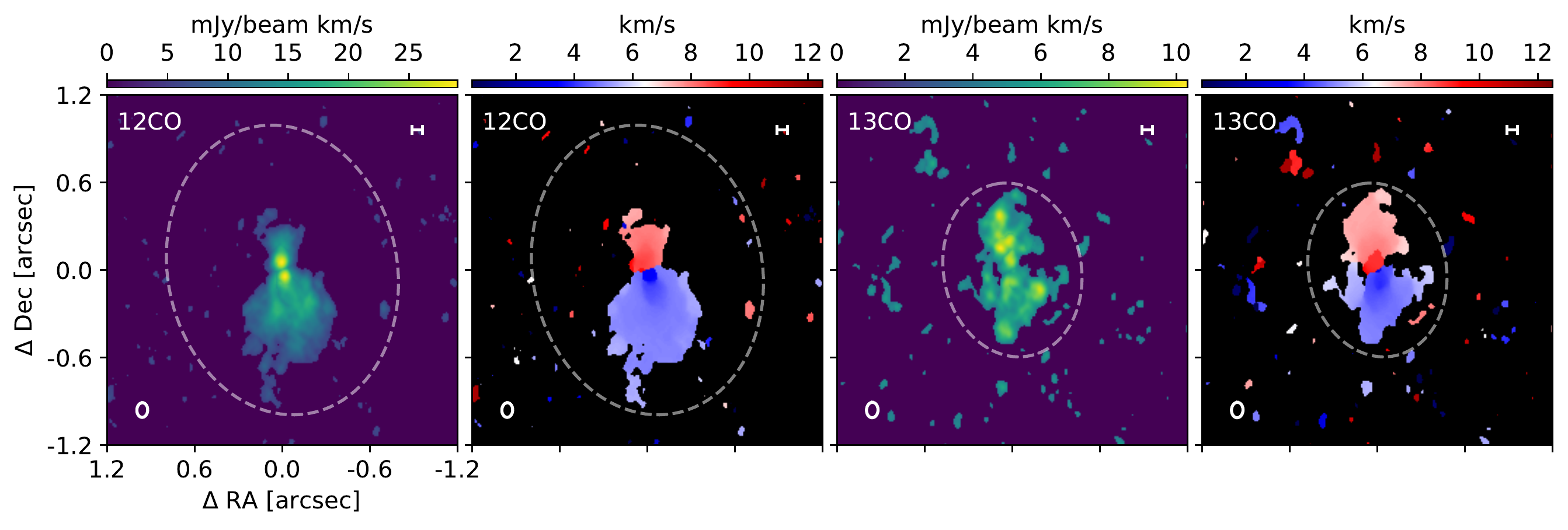}
   \caption{Moment 0 and moment 1 of the $^{12}$CO and $^{13}$CO. The upper right bar corresponds to 10\,au size. The dashed circles (projected with the same inclination and PA as CIDA 1, Table~\ref{tab:mcmc_results}) are 1.0'' and 0.6'' in size  from the center for the $^{12}$CO and $^{13}$CO, respectively. The details of the images are summarized in Table~\ref{tab:summary_ALMA}.}
   \label{fig:CIDA_gas}
\end{figure*}

Nevertheless, in previous ALMA observations the cavity of CIDA1 was marginally resolved and the nature of this structure was unclear. It therefore remains unclear as to whether or not the cavity is empty of millimeter-sized particles, whether or not the ring is composed of multiple structures, and whether or not there are variations of the spectral index throughout the disk that indicate grain growth in particular regions. In this paper, we present observations with ALMA of CIDA\,1 in Band 7 (0.9\,mm) and Band\,4 (2.1\,mm) with a high angular resolution at the two wavelengths ($\sim 0.05''\times 0.034''$) in order to potentially resolve small-scale features and resolve possible variations of the spectral index throughout the disk. In addition, X-shooter observations on the Very Large Telescope (VLT) were acquired to get better constraints of the CIDA\,1 accretion rate, which can help to test models of cavity formation, such as photoevaportation \citep[e.g.,][]{ercolano2017}. This paper is organized as follows. Section~\ref{sect:observations} summarizes the observations and data reduction. Section~\ref{sect:results} presents the results of our X-Shooter and ALMA observations, and the analysis of the dust morphology in the visibility plane. Section~\ref{sect:models_obs} discusses our observations in the context of dust evolution models and planet--disk interaction. Finally, in Sect.~\ref{sect:discussion} we discuss the results of our observations, and in Sect.~\ref{sect:conclusions} we present our conclusions. 

\begin{figure*}
 \centering
 \tabcolsep=0.05cm 
   \begin{tabular}{ccc}   
        \includegraphics[width=6.0cm]{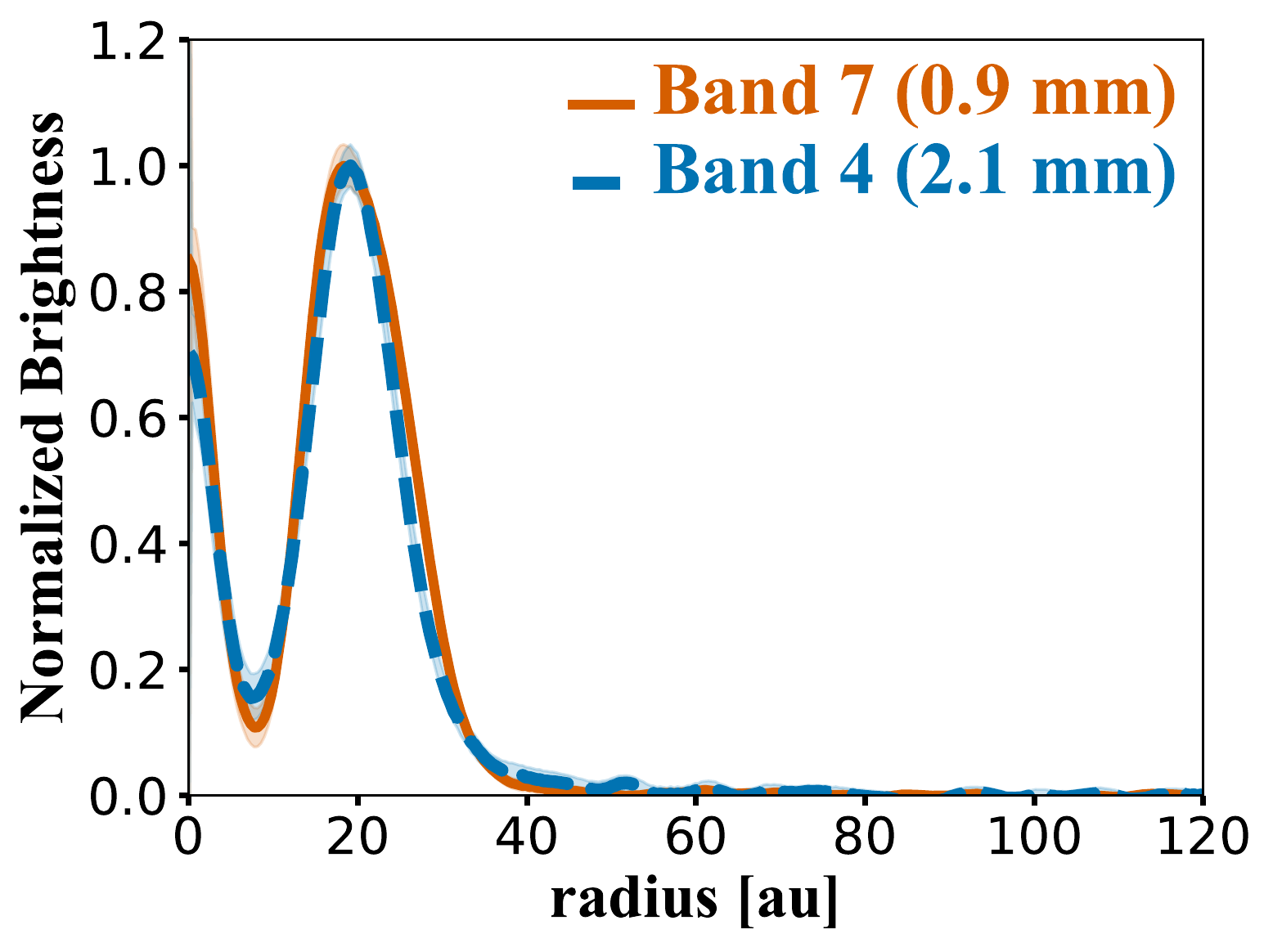}&
        \includegraphics[width=6.2cm]{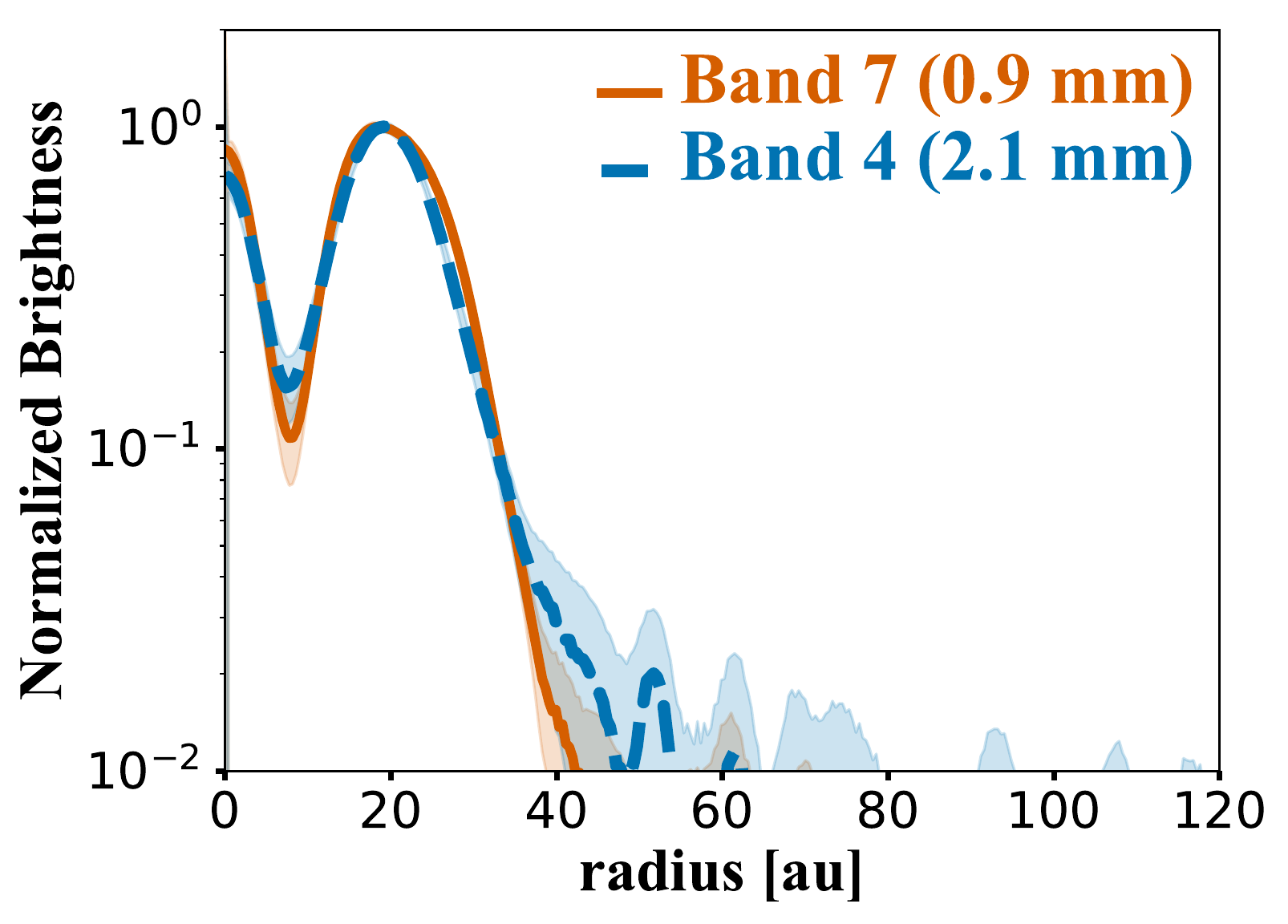}&
        \includegraphics[width=6.0cm]{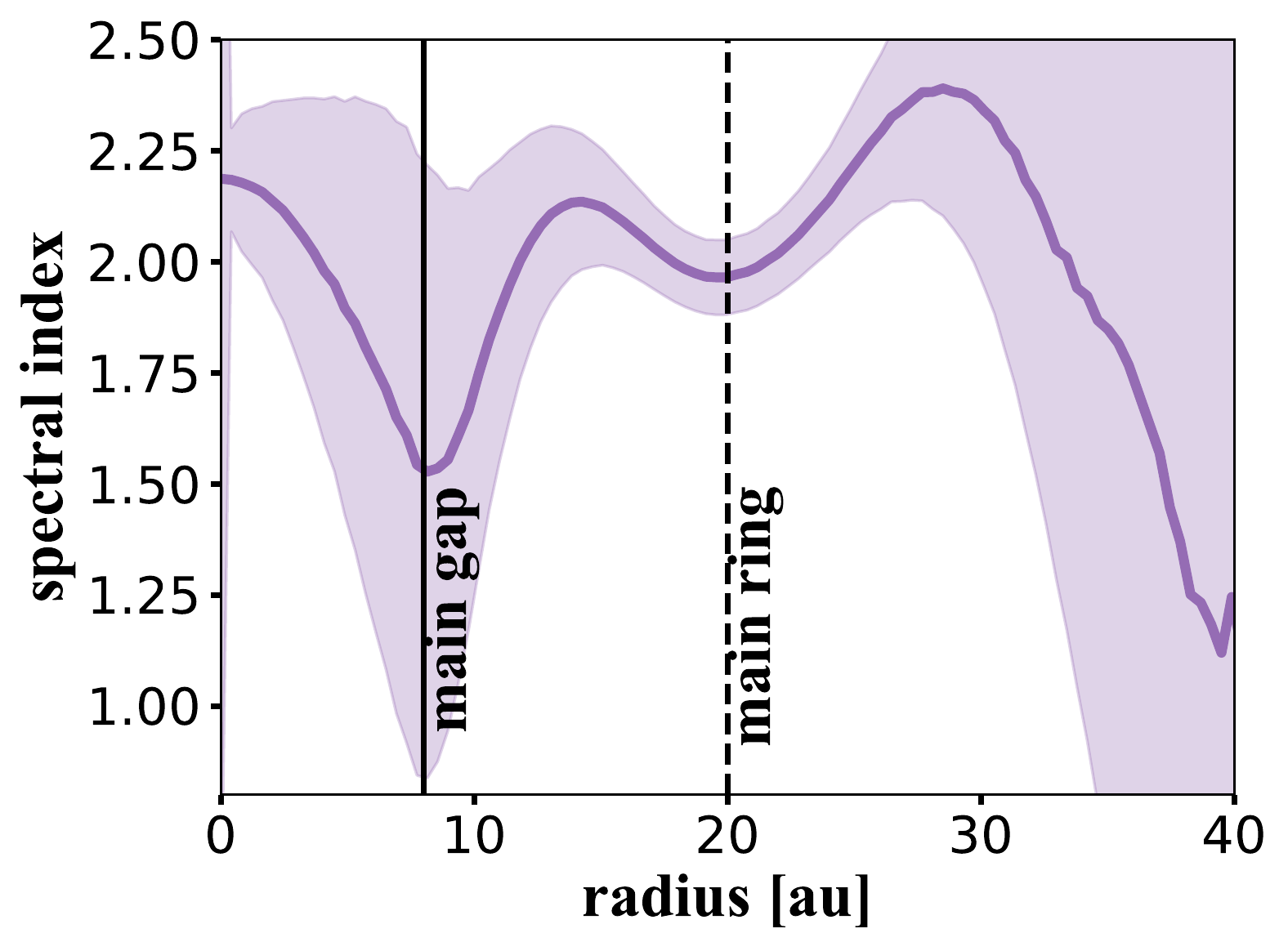}
    \end{tabular}
   \caption{Azimuthally averaged radial intensity profiles of the deprojected continuum images (lineal scale on the left and logarithmic scale in the center) at each wavelength. Each profile is normalized to the peak. The profiles are obtained from images that are restored with the same beam (50mas $\times$ 34mas). The error bars include the standard deviation of each elliptical bin divided by the square root of the number of beams spanning the full azimuthal angle at each radial bin. The right panel shows the radial profile of the spectral index taking into account the errors on the radial intensity profiles in addition to 10\% of the error due to flux calibration.}
   \label{fig:radial_profile}
\end{figure*}

\section{Observations}  \label{sect:observations}

\begin{table*}
\begin{center}
\caption{Summary of ALMA observations. \label{tab:obs_log}}
\begin{tabular}{ |l|c|c|c|c|c|c| } 
    \hline
    \hline
    Band    & Program ID     &  Obs. Date  & Exp. time & N$^\circ$ & Baselines & Configuration \\
            &                &             & (min)     & Antennas  & (m)       &               \\
    \hline
    4       & 2018.1.00536.S &  2018-10-19 & 5.99      & 44        & 15 - 2516 & Compact   \\
            &                &  2019-06-09 & 32.51     & 41        & 83 - 16196& Extended  \\
            &                &  2019-06-23 & 37.63     & 45        & 83 - 16196& Extended  \\
            &                &  2019-07-04 & 37.95     & 46        & 83 - 16196& Extended  \\
            &                &  2019-07-04 & 38.10     & 46        & 83 - 16196& Extended  \\
            &                &  2019-07-15 & 24.14     & 41        & 139 - 8548& Compact   \\
    \hline
    7       & 2015.1.00934.S &  2016-08-12 & 47.68     & 38        & 15 - 1462 & Compact   \\
            & 2016.1.01511.S &  2017-07-06 &  4.23     & 42        & 17 - 2647 & Compact   \\
            & 2018.1.00536.S &  2019-07-19 & 36.24     & 47        & 92 - 8548 & Extended  \\
            & 2018.1.00536.S &  2019-07-19 & 35.92     & 45        & 92 - 8548 & Extended  \\
    \hline
    \hline
\end{tabular}
\end{center}
\end{table*}

\subsection{ALMA observations}

The data sets studied in this work include ALMA observations at high angular resolution ($\approx0.05\times0.034''$) of CIDA1 at $0.9$\,mm and $2.1$\,mm wavelengths, under the ALMA project 2018.1.00536.S (PI: A.~Natta). The $0.9\,$mm data were obtained by ALMA Band\,7, and the correlator was configured to observe four spectral windows: two covered dust continuum emission centered at $344.0\,$GHz and $333.25\,$GHz, and two  centered at the molecular lines $^{12}$CO ($J:3-2$) and $^{13}$CO ($J:3-2$). The frequency resolution of the  channels is $976.562\,$kHz for continuum and $^{13}$CO spectral windows, while it was $244.141\,$kHz  for $^{12}$CO  (approximately $1\,$km\,s$^{-1}$ and $0.2\,$km\,s$^{-1}$, respectively). For the imaging and analysis, we combined all available ALMA Band\,7 observations, thus including the archival data from 2015.1.00934.S (PI: L.~Ricci) and 2016.1.01511.S (PI: J.~Patience) already published in \cite{pinilla2018} and \cite{kurtovic2021}.

Observations at $2.1\,$mm were carried out with ALMA Band\,4. The correlator was configured to observe four spectral windows: three were located to observe dust continuum emission, centered at $134.4\,$GHz, $136.2\,$GHz, and $148.2\,$GHz, while the remaining  one was centered at the molecular line CS $J:3-2$. The frequency resolution of the continuum channels  was $976.562\,$kHz, while the CS line was observed at $31\,$kHz (approximately $2.4\,$km\,s$^{-1}$ and $0.1\,$km\,s$^{-1}$, respectively). The details of all the data used in this paper are summarized in Table~\ref{tab:obs_log}.

After ALMA standard pipeline calibration, a series of preparation steps have to be applied to combine and self-calibrate the datasets. Using \texttt{CASA 5.6.2}, we extract the dust continuum emission by flagging the channels located at $\pm 25\,$km\,s$^{-1}$ from each targeted spectral line. The remaining channels from all spectral windows are averaged into 125\,MHz channels. To avoid differences in the weight calculation through different ALMA cycles, we apply the task \texttt{statwt} to recalculate the visibility weighting according to the observed scatter. Next, we aligned the datasets and compared the flux calibration of each individual execution to ensure they were consistent. In Band\,7, we found and corrected a 10\% discrepancy between the observation from project 2015.1.00934.S and all the other observations, and so we rescaled this compact configuration data to match the others. In Band\,4, we corrected a discrepancy of $19\%$ between the first observation of the extended baselines  and all others.

To enhance the signal-to-noise ratio (S/N), self-calibration was performed in two stages. First, we self-calibrated the data sets we classified as having ``compact'' baseline configuration in Table\ref{tab:obs_log}, and then we combined these with the ``extended'' baselines and self-calibrated again. We used a Briggs robust parameter of 0.5 for the imaging of the self-calibration process. The combined data sets from Band\,4 had a peak S/N of $\approx23$ before self-calibration, thus we only applied one phase and one amplitude calibration, using the whole integration time as the solution interval. 

In order to reduce the data volume for the visibility analysis, we averaged the continuum emission into channels of 250\,MHz in width  and 30\,s time-binning. We used the central frequency of each binned channel  to convert the uv-coordinates into wavelength units.

All the dust continuum calibration steps, including the centroid shifting, flux calibration, and self-calibration tables, were then applied to the molecular line emission channels. The continuum emission was subtracted using the \texttt{uvcontsub} task. We generate images of $0.5$\,km\,s$^{-1}$ in velocity width, with a robust parameter of 1. To enhance the S/N, we chose to apply a uv-tapering with a Gaussian of $0.03''$ on the $^{12}$CO, and $0.04''$ on the $^{13}$CO. We used the package \texttt{bettermoments} \citep{bettermoments} to create the kinematic map. This package fits a quadratic function to find the peak intensity of the line emission in each pixel, and the velocity associated to it.

\subsection{X-Shooter observations}

We acquired one broad-band flux-calibrated spectrum of CIDA\,1 with the X-Shooter instrument on the VLT. The observations were taken for Pr.Id. 105.2061.001 (PI: A.~Natta), on the night of October 30, 2020. Sky conditions were clear (CLR), seeing at zenith was $\sim$1\arcsec. The spectrum was observed using slit width 1.0\arcsec-0.4\arcsec-0.4\arcsec ~ in the three arms (UVB, VIS, NIR) of the X-Shooter spectrograph, respectively. These spectra were reduced using the X-Shooter pipeline \citep{modigliani11} v3.5.0 run through the ESOReflex tool. This pipeline performs the main reduction steps, including flat fielding, correction for bias and dark, wavelength calibration, rectification, and extraction of the 1D spectrum. The latter is then corrected for telluric absorption lines using the molecfit tool, version 3.0.3 \citep{molecfit}. The absolute flux calibration of the spectra is then performed by scaling the spectra to the one obtained in the same observing block using a wider set of 5.0\arcsec ~ slits. The flux rescaling factors are ~1.2, 3.7, and 3.5 in the three arms, in line with the fact that the slit widths in the three arms are smaller than the seeing at the corresponding wavelengths.

\section{Results from ALMA and X-Shooter observations} \label{sect:results}

\begin{table*}
\begin{center}
\caption{Summary of the ALMA images in Fig.~\ref{fig:ALMA_CIDA}.} \label{tab:summary_ALMA}
\begin{tabular}{|c|c|c|c|c|c|} 
    \hline
    \hline
    Band & $\lambda$ & $F_{\rm{peak}}$ & $F_{\rm{total}}$ & $\sigma$ & beam\\
         &[mm]&[Jy\,beam$^{-1}$]& [Jy]& [Jy\,beam$^{-1}$] &[mas$\times$mas]\\
    \hline     
    7&0.9&$7.9\times10^{-4}$&$3.1\times10^{-2}$&$1.7\times10^{-5}$&54$\times$34\\
    \hline
    4&2.1&$1.4\times10^{-4}$&$5.1\times10^{-3}$&$5.7\times10^{-6}$&50$\times$34\\
    \hline
    \hline
\end{tabular}
\end{center}
\end{table*}

\subsection{Images, dust disk mass, and spectral index}\label{sect:images}

Figure~\ref{fig:ALMA_CIDA} shows the final images of CIDA\,1 of the ALMA observations in Band\,7 (0.9\,mm) and in Band\,4 (2.1\,mm) and the details of the images are summarized in Table~\ref{tab:summary_ALMA}. The fluxes from this table are obtained from the images assuming all the emission within a circular area of 0.5'' from the center. The images of the $^{12}$CO and $^{13}$CO are shown in Figure~\ref{fig:CIDA_gas}; these were obtained by processing the channel maps with the Python package \texttt{bettermoments} \citep{bettermoments}, and clipping the emission at 3$\sigma$. The emission of CS was not detected in Band 4, although an increase in the noise level is observed close to the central velocity channel. We imaged the channels around the expected location of the CS line with a velocity resolution of $1.0\,$km\,s$^{-1}$. Three moment 0 images were generated from these channel maps, spanning the velocities from 1-4, 5-8, and 9-12\,km\,s$^{-1}$. When the rms is measured in a circle of $1\arcsec$ centered on CIDA\,1, the moment 0 image spanning 5-8km\,s$^{-1}$ has a rms of 1.56 mJy\,beam$^{-1}$, while the other two moment 0 images have an rms of 1.04 mJy\,beam$^{-1}$. 

The dust continuum images at both wavelengths reveal unresolved emission from the inner disk, a large gap, and a ring-like structure peaking at around 20\,au. We use the distance to CIDA\,1 of 137.5$\pm$1.1\,pc derived by \cite{bailer-jones2018} using Gaia DR2 \citep{gaia2018}. Figure~\ref{fig:radial_profile} shows in linear and logarithmic scale the azimuthally averaged radial intensity profiles of the deprojected continuum images at each wavelength (using the inclination and position angle report in Table~\ref{tab:mcmc_results}). For this figure, each profile is normalized to the peak of the ring and the images are previously restored with the same beam before deprojection (0.05\arcsec $\times$ 0.034\arcsec). For the deprojection, we use the inclination (inc) and position angle (PA) from Table~\ref{tab:mcmc_results}, which are obtained from the visibility analysis explained in Sect.~\ref{sect_visi_analysis}. The error bars include the standard deviation of each elliptical bin divided by the square root of the number of beams spanning full azimuthal angle at each radial bin. In this figure, it is possible to see that emission at both wavelengths peaks at the ring location (20\,au) and there is a very bright inner disk. The $^{12}$CO peak intensity map also shows a cavity that remains unresolved with a size of $\sim7$\,au along the major axis of the disk. Another interesting result from  the $^{12}$CO peak intensity map is that this emission is more extended than the dust continuum emission, as found in \cite{kurtovic2021}.  The analysis of the $^{12}$CO and $^{13}$CO will be presented in a future paper by Curone et al. (in prep).

\begin{figure*}
    \centering
    \includegraphics[width=18.0cm]{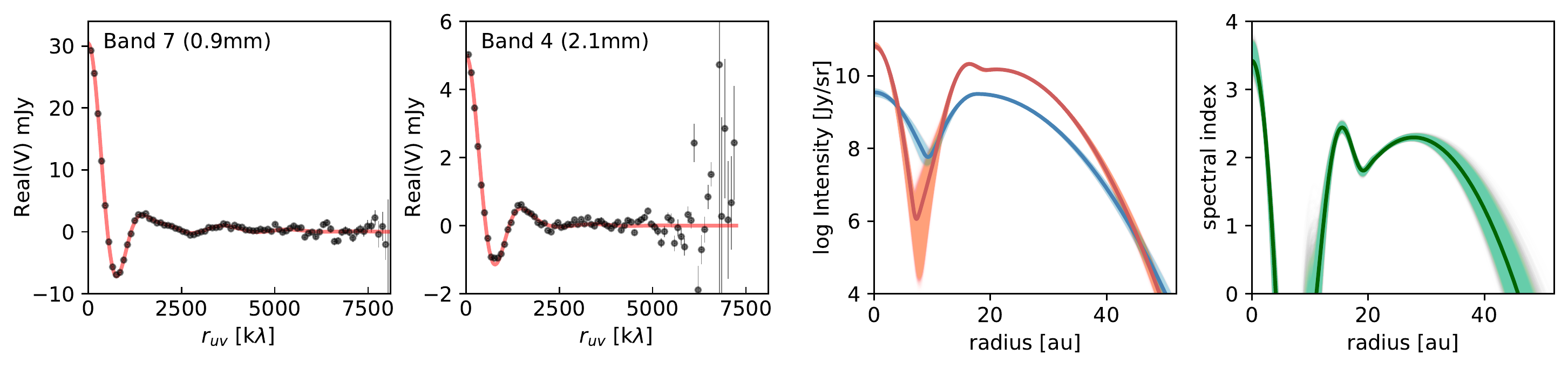}
    \caption{From left to right: Real part of the visibilities after centering and deprojecting the data in Band 7 and in Band 4 vs. the best-fit model from \texttt{galario} of the continuum data. Intensity profile from the best fit in Band 7 and Band 4 (orange and blue line, respectively), with  5000 randomly selected chains after convergence overlaid. The right panel shows the spectral index radial profile obtained from the intensity profiles.}
    \label{fig:bestfit}
\end{figure*}

\begin{figure*}
    \centering
    \includegraphics[width=18.0cm]{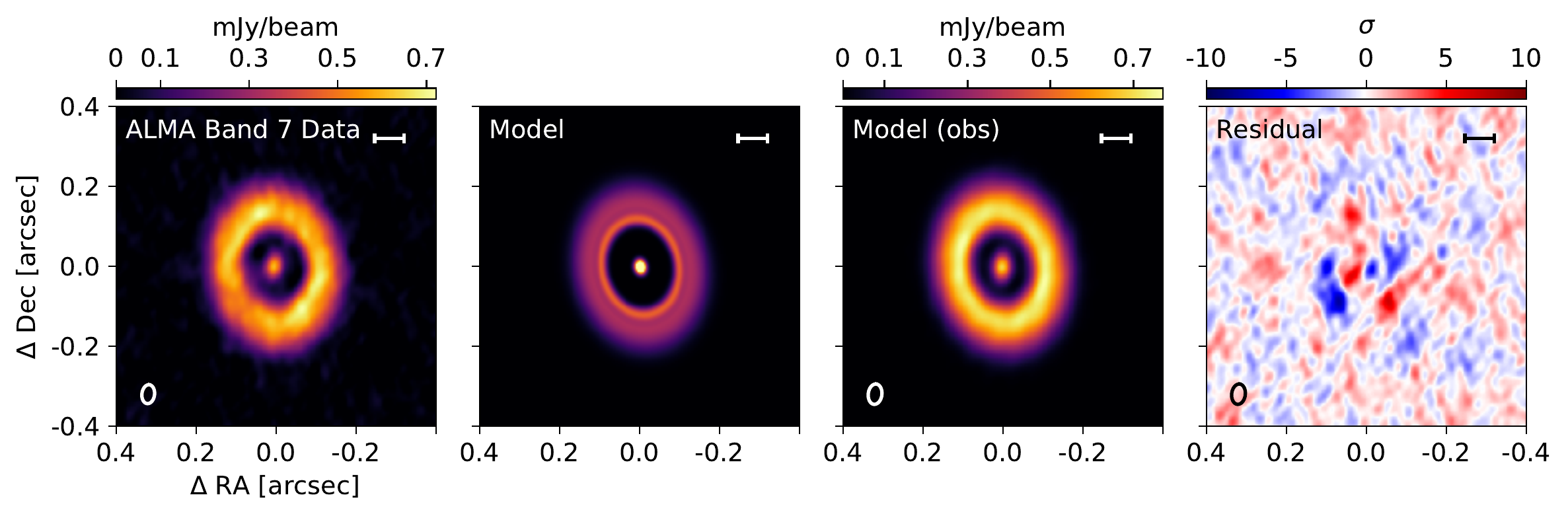}
    \includegraphics[width=18.0cm]{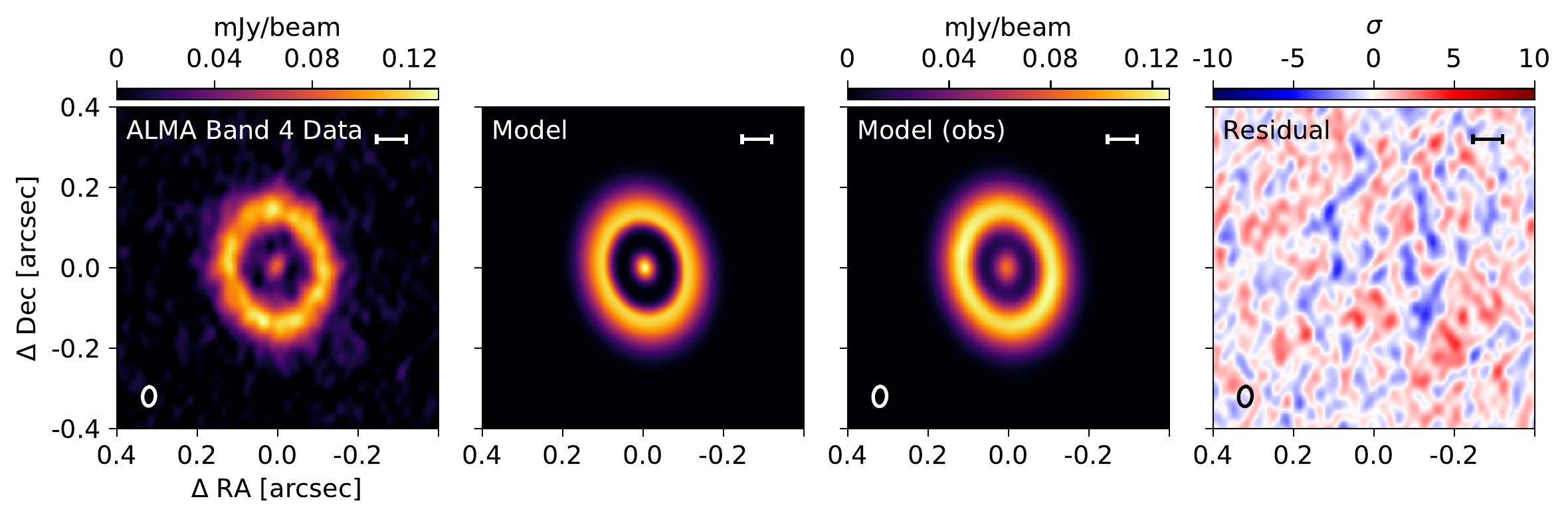}
    \caption{Observations vs. model images from our best fit with \texttt{galario} (before and after convolution) and the residuals. Top panels correspond to Band 7 and the bottom panels correspond to Band 4.}
    \label{fig:Galario_Band47}
\end{figure*}

We calculate the optical depth of the peak of the continuum ring, assuming $\tau=-\ln[1-T_{\rm{brightness}}/T_{\rm{physical}}]$, with $T_{\rm{brightness}}$ and $T_{\rm{physical}}$ being the brightness and physical temperature respectively. We obtain the brightness temperature from the full black-body Planck function. We used a radially constant physical temperature of  20\,K and we find that the optical depth at the peak is $\tau_{\rm{peak}, B7}=0.48$ and $\tau_{\rm{peak}, B4}=0.28$ for Band 7 and Band 4, respectively. We also calculate $\tau$ assuming a luminosity-dependent relation for the temperature from \cite{plas2016}, that is, $T=22(L_\star/L_\odot)^{0.16}$K, in which case we obtain a physical temperature of 14.7\,K and $\tau_{\rm{peak}, B7}=0.74$ and $\tau_{\rm{peak}, B4}=0.41$. Using our radiative transfer calculations presented in Sect.~\ref{sect:models_obs} for CIDA\,1, the temperature expected at the midplane and at the peak location ($\sim$20\,au) is around 30\,K, in which case $\tau_{\rm{peak}, B7}=0.29$ and $\tau_{\rm{peak}, B4}=0.18$.  We therefore assume that most of the emission is optically thin to calculate the dust disk mass. These values for the optical depths fall in the same range as reported by the DSHARP sample \citep{dullemond2018} for more massive disks. One potential explanation for this range of the optical depth inside the ring is planetesimal formation in this region \citep[][]{stammler2019}.

Under the assumption of optically thin emission, we calculate the dust disk mass as  $M_{\mathrm{dust}}\simeq\frac{{d^2 F_\nu}}{\kappa_\nu B_\nu (T)}$, where $d$ is the distance to the source, $F_\nu$ is the total flux at a given wavelength, and $B_\nu$ is the blackbody surface brightness at a given temperature \citep{hildebrand1983}. Taking a mass absorption coefficient ($\kappa_\nu$) at a given frequency as $\kappa_\nu=2.3\,$cm$^{2}$\,g$^{-1}\times(\nu/230\,\rm{GHz})^{0.4}$ \citep[which is typical for a population of large millimeter grains,][which is compatible within 2$\sigma$ with the spectral index we infer for CIDA\,1 as shown later]{beckwith1990, andrews2013}, and a dust temperature of 20.0\,K or 14.7\,K, the dust disk mass is 7.4\,$M_{\rm{Earth}}$ or 12.0\,$M_{\rm{Earth}}$, respectively, when taking the total flux from the image in Band 7. For Band 4, the dust disk mass is 7.8 \,$M_{\rm{Earth}}$ or 11.4\,$M_{\rm{Earth}}$ for 20\, or 14.7\,K, respectively. Therefore, the inferred dust disk mass is consistent between the two wavelengths with a large uncertainty that depends on the assumed temperature. In addition, $\kappa_\nu$ is a very unknown quantity in protoplanetary disks, and for direct comparison we take the same $\kappa_\nu$ assumed to calculate the dust disk mass of large populations of protoplanetary disks \citep[e.g.,][]{pascucci2016}. Recent work from \cite{lin2020} about inferring (sub)millimeter dust opacities from resolved multi-wavelength continuum observations of edge-on disks provides support to this widely used opacity assumption from \cite{beckwith1990}. 

With the total flux from the images at the two wavelengths, we obtained the spatially integrated spectral index as $\alpha_{\mathrm{mm}}$=$\log(F_{\rm{B7}}/F_{\rm{B4}})/\log(\nu_7/\nu_4)=2.0\pm0.2$, where the uncertainty includes the rms of the observations and 10\% uncertainty from the flux calibration. This value is in agreement with the values found in \cite{ricci2014}. The very low fluxes within the gap do not allow us to obtain reliable values of the spectral index in this region (see right panel of Fig.~\ref{fig:radial_profile}). When taking the emission either from a circled area that is centered at the location of CIDA\,1 and that is as big as the beam or an area that encircles the ring, the integrated spectral index in each region is around $2.0$. The right panel of Fig.~\ref{fig:radial_profile} shows the radial profile of the spectral index, which shows a slight decrease inside the ring, but the differences are not significant ($<1\sigma$). Within the uncertainties, the spectral index seems to be relatively constant throughout the disk, as found by \cite{tazzari2020} in a sample of disks in the Lupus star formation region.

\begin{table}
\centering
\caption{Best parameters from the uv modeling, following equations~\ref{eq:1g1bg_param_model} and \ref{eq:2g1bg_param_model}. Here, ``mas'' stands for milliarcsecond.}
\begin{tabular}{ c|c|c|c } 
  \hline
  \hline
\noalign{\smallskip}
    & Band 4  & Band 7  & units \\
    & (2.1mm) & (0.9mm) &       \\
\noalign{\smallskip}
  \hline
\noalign{\smallskip}
    $\delta_{\rm{RA}}$  & $ 3.5_{-0.2}^{+0.2}$ & $ 1.2_{-0.1}^{+0.1}$ & mas \\
    $\delta_{\rm{Dec}}$ & $-1.5_{-0.3}^{+0.2}$ & $-3.4_{-0.2}^{+0.2}$ & mas \\
    inc                 & $37.7_{-0.3}^{+0.1}$ & $37.4_{-0.1}^{+0.1}$ & deg \\
    PA                  & $12.7_{-0.4}^{+0.3}$ & $10.8_{-0.1}^{+0.2}$ & deg \\
\noalign{\smallskip}
  \hline
\noalign{\smallskip}
    $\log f_0$       & $9.54_{-0.04}^{+0.02}$ & $10.82_{-0.01}^{+0.04}$ & $\log_{10}$(Jy/sr) \\
    $\sigma_0$    & $22.2_{-0.8}^{+1.4}$   & $10.9_{-0.6}^{+0.1}$ & mas \\
\noalign{\smallskip}
    $\log f_1$         & $9.50_{-0.01}^{+0.01}$ & $10.18_{-0.01}^{+0.01}$ & $\log_{10}$(Jy/sr) \\
    $r_1$         & $130.6_{-1.5}^{+0.4}$  & $155.7_{-1.7}^{+0.8}$ & mas \\
    $\sigma_{1i}$ & $20.2_{-1.3}^{+0.4}$   & $23.0_{-5.4}^{+0.7}$ & mas \\
    $\sigma_{1o}$ & $47.5_{-0.4}^{+0.8}$   & $38.7_{-0.5}^{+0.8}$ & mas \\
\noalign{\smallskip}
    $\log f_2$         & ---                    & $10.22_{-0.01}^{+0.06}$ & $\log_{10}$(Jy/sr) \\
    $r_2$         & ---                    & $118.6_{-0.5}^{+0.8}$ & mas \\
    $\sigma_2$    & ---                    & $11.3_{-0.5}^{+0.6}$ & mas \\
\noalign{\smallskip}
  \hline
\noalign{\smallskip}
    $R_{68}$      & $23.6\pm0.1$ & $24.1\pm0.1$ & mas \\
    $R_{95}$      & $28.4\pm0.1$ & $28.6\pm0.1$ & mas \\
    $F_\lambda$   & $4.9\pm0.1$ & $29.2\pm0.1$ & mJy \\
\noalign{\smallskip}
  \hline
  \hline
\end{tabular}
\label{tab:mcmc_results}  
\end{table}

\subsection{Dust morphology from visibility fitting} \label{sect_visi_analysis}

We describe the emission on both ALMA bands with axisymmetric parametric models, and for comparison we also use the Gaussian Processes modeler Python package \texttt{frank} \citep{jennings2020}. While a single centrally peaked Gaussian is enough to describe the inner emission of CIDA\,1, for the ring emission we tried several different combinations of Gaussian profiles and/or radially broken Gaussian profiles (Gaussians with different widths to each side of the peak). After these tests, we find that the best parametric models for the ALMA Band 4 (2.1mm) and Band 7 (0.9mm) emission are given by:

\begin{equation}
    I_{\rm{2.1mm}}(r) = 10^{f_0}\,g(r, \,\sigma_0) + 10^{f_1}\,g_{\rm{bg}}(r-r_1, \,\sigma_{1i}, \,\sigma_{1o}) \text{,}
    \label{eq:1g1bg_param_model}
\end{equation}

\begin{equation}
    \begin{aligned}
        I_{\rm{0.9mm}}(r) = \,\,\,\,\, & 10^{f_0}\,g(r, \,\sigma_0) + 10^{f_1}\,g_{\rm{bg}}(r-r_1, \,\sigma_{1i}, \,\sigma_{1o}) \\
        + & 10^{f_2}\,g(r, \,\sigma_2) \text{,}
    \end{aligned}
    \label{eq:2g1bg_param_model}
\end{equation}

\noindent where $g$ is a Gaussian profile, and $g_{\rm{bg}}$ is a broken Gaussian, described by:
\begin{equation}
    g(x, \,\sigma) = \exp{\left( - \frac{x^2}{2\,\sigma^2} \right)} \text{,}
    \label{eq:gaussian}
\end{equation}

\begin{equation}
    g_{\rm{bg}}(x, \,\sigma) = \left\{
    \begin{aligned}
        g(x, \sigma_i) \,\,\,\, \text{for }x \leq 0\text{.}\\
        g(x, \sigma_o) \,\,\,\, \text{for }x> 0\text{.}
    \end{aligned} \right.
    \text{}
    \label{eq:broken_gaussian}
\end{equation}

For each model, the visibilities are obtained by optimizing the model profile with a spatial offset ($\delta_{\rm{RA}}$, $\delta_{\rm{Dec}}$), an inclination and PA, which are used to deproject the observational data. Therefore, each model has four extra free parameters in addition to those that describe the intensity profile. The Fourier Transforms to obtain the models visibilities and the $\chi^2$ calculation are computed with the \texttt{galario} Python package \citep{tazzari2018} using a pixel size of 0.2\,mas.

We sample the posterior probability distribution with a Markov chain Monte Carlo (MCMC) routine based on the \texttt{emcee} Python package \citep{emcee}. We used a flat prior probability distribution over a wide parameter range, such that the walkers would only be initially restricted by geometric considerations (inc $\in [0,90]$ , PA $\in [0,180]$, $\sigma \geq 0$).
We ran more than 250000 steps after convergence to find the most likely parameter set for each model, as well as taking the 16th and 84th percentile for the error bars. Our results are shown in Table \ref{tab:mcmc_results}.

Figure~\ref{fig:bestfit} shows the real part of the visibilities after centering and deprojecting the data in Band 7 and in Band 4 vs. the best fit model that minimizes $\chi^2$ from \texttt{galario} of the continuum data. In addition, this figure includes the intensity profile from the best fit in Band 7 and Band 4 (orange and blue line, respectively), with the 5000 randomly selected chains after convergence  overlaid, and  the right panel shows the spectral index radial profile obtained from the intensity profiles. The intensity profiles from the best fit reveal that the ring in Band\,7 seems to have a shoulder after the peak of emission that is not visible in the Band\,4 data. It is worth noting that a fit in Band 7 without this additional ring in the outer disk (e.g., the model used for Band 4) produced significant residuals, suggesting the presence of a further ring.  In addition, our best fits show that the inner disk is as bright as or brighter than the peak brightness of the ring. The gap seems to be shallower in Band 4, but this result needs to be taken with care because the S/N of Band 4 is around half that of Band 7, and to confirm these differences between Band 7 and Band 4, deeper observations are required in Band 4. The residuals of the Band 7 fit hint at a nonaxisymmetric structure of the ring, and deeper observations at this wavelength are also needed to check for this potential asymmetry.

From the best intensity profiles, we calculate the radial profile of the spectral index presented in the right panel of Fig.~\ref{fig:bestfit}, which shows the same trend as those seen in the images (right panel of Fig.~\ref{fig:radial_profile}), where there is a hint of a slight decrease inside the ring and overall a low value of the spectral index in the entire disk ($\sim$2) with a high decrease in the gap. However, the spectral index there is not trustworthy because of very low fluxes and large uncertainties as shown in the right panel of Fig.~\ref{fig:radial_profile}. We also check if there is any evidence from the visibilities that the spectral index varies by taking the ratio of the visibilities, and this ratio is  constant within the uncertainties.

Figure~\ref{fig:Galario_Band47} shows the observations versus model images from our best fit with \texttt{galario} (before and after convolution) and the residuals, for both Band 7 and Band 4. Our residuals do not show any clear sign of additional substructures, either radial or azimuthal at the current resolution and sensitivity of the observations.

\subsection{Stellar properties from the X-shooter spectrum} \label{sect_stellar_properties}

\begin{figure}
    \centering
    \includegraphics[width=9.0cm]{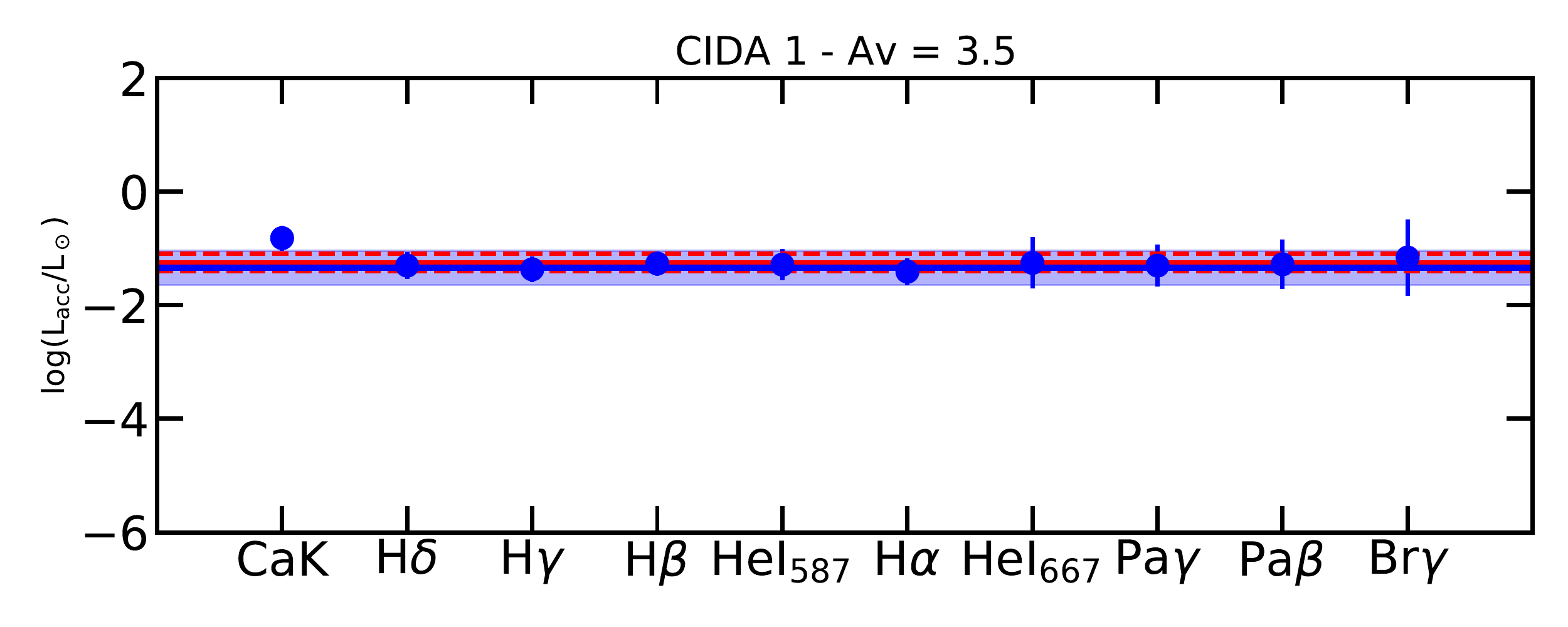}
    \caption{Accretion luminosity derived from the luminosity of various emission lines, as labeled, using the relations by \citet{alcala17}, and assuming $A_V=3.5$ mag. The blue line is the best-fit accretion luminosity obtained fitting the whole X-Shooter spectrum of CIDA\,1. The red line is the mean value of $L_{\rm acc}$ obtained from the luminosity of the emission lines.}
    \label{fig:lacc_lines}
\end{figure}

The spectrum of CIDA~1 is rich in permitted emission lines and shows a prominent Balmer jump. We fitted the spectrum following the procedure described by \citet{manara13}. Briefly, we explore a parameter space including several photospheric templates of young stars, a set of slab models to reproduce the accretion emission continuum spectra, and increasing extinction value until the sum of the photospheric template and the accretion spectrum better reproduce the de-reddened spectrum of CIDA\,1. In this case, the solution is highly degenerate between templates with spectral types between M4.5 and M6.5, and extinction values $A_V\sim2-4$ mag. We therefore use an independent method to constrain the best values for $A_V$, and we measure the emission line fluxes for a number of permitted emission lines in the whole X-Shooter spectrum. We then convert the measured line luminosity in accretion luminosity using the relations by \citet{alcala17}, and repeat this exercise by increasing the amount of $A_V$ from 0 to 4 mag in steps of 0.25 mag. The accretion luminosity derived from the lines shows the minimum dispersion with $A_V\sim3.5$ mag (see Fig.~\ref{fig:lacc_lines}). 

\begin{figure}
    \centering
    \includegraphics[width=9.0cm]{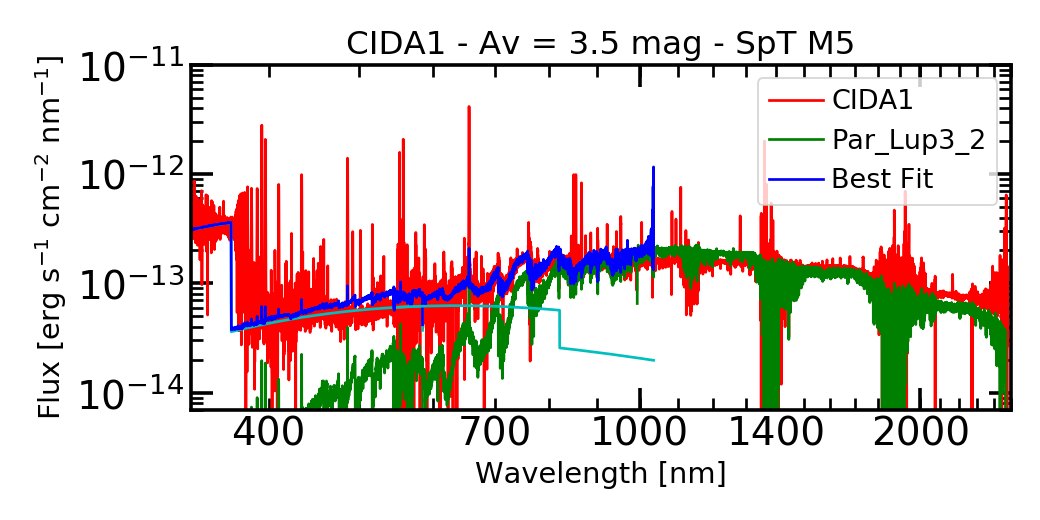}
    \caption{Fit of the spectrum of CIDA\,1 obtained here. The dereddened spectrum of the target (red) is reproduced by the sum of a photospheric template (green) and a slab model to reproduce the accretion shock emission (cyan). }
    \label{fig:lacc_lines2}
\end{figure}

Constraining the values of $A_V$ between 3 and 3.8 mag, we obtain a best fit from the X-Shooter spectrum following the evolutionary tracks from \cite{baraffe2015}, with SpT M5, $A_V$= 3.5 mag, resulting in a value of stellar luminosity $L_\star$ = 0.15 $\pm$ 0.03 $L_\odot$, accretion luminosity $\log (L_{\rm acc}/L_\odot)$ = $-$1.34 $\pm$ 0.3, a stellar mass $M_\star$ = 0.1-0.2 $M_\odot$, and a mass accretion rate $\dot{M}_{\rm acc}$ = 1.4 $~\times~10^{-8} M_\odot$/yr. The best fit is shown in Fig.~\ref{fig:lacc_lines2}. This value of mass accretion rate is in line with the results of \citet{HH08}, and is higher than the one we reported in \citet{pinilla2018}. The addition of the infrared emission lines in the analysis allows us to better constrain the extinction value from the emission lines. 

We note that CIDA\,1, despite being a strong accretor, shows very little emission in forbidden lines, with only the [OI]$\lambda$6300\AA ~ line detected in the spectrum. The [OI] emission may be a tracer of some winds in the inner part of the disk \citep{simon2016, mcginnis2018, banzatti2019}. These authors find that [OI] lines in transition disks are narrow, like the one we see in CIDA\,1,  but it is unclear in this case whether it originates from photoevaporation \citep[e.g.,][]{ercolano2016} or a magneto-hydrodynamical (MHD) wind \citep[e.g.,][]{banzatti2019}.

\section{Comparison with dust evolution models} \label{sect:models_obs}

In this section, we compare our observations with dust evolution models in combination with radiative transfer calculations. Our models are in the context of giant planets opening a gap in the disk and potentially creating a cavity to explain the observed properties of transition disks.

\begin{figure*}
    \centering
    \includegraphics[width=18.0cm]{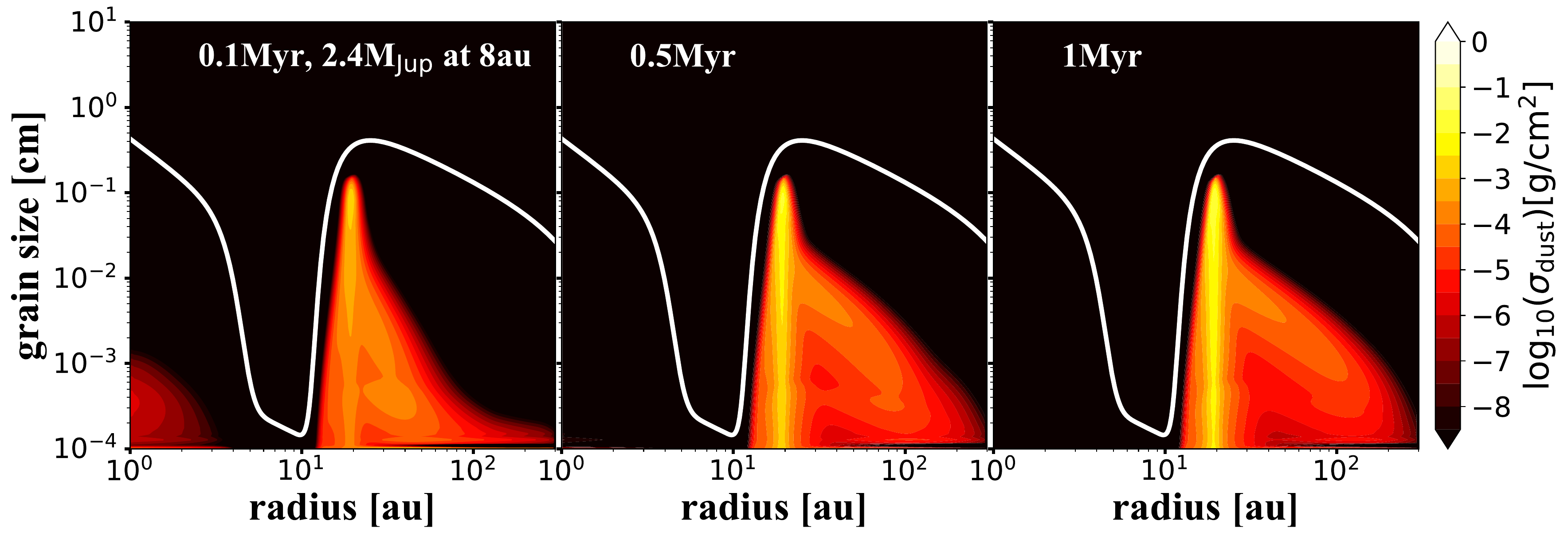}
    \caption{Particle density distribution after 0.1\,Myr, 0.5\,Myr, and 1\,Myr of evolution as a function of the distance from the star and grain size when a $\sim$2.4\,$M_{\rm{Jup}}$  planet is located at 8\,au from the star. }
    \label{fig:dust_evo2}
\end{figure*}

A pressure bump at the outer edge of a planetary gap can trap millimeter-sized particles and can create a dust-depleted cavity as seen at millimeter wavelength depending on the mass of the planet. The planet mass also determines whether or not an inner disk survives after million-year timescales. In a subset of transition disks, near-infrared (NIR) excess has been detected, which is interpreted as an optically thick dusty inner disk located within the first astronomical units \citep[e.g.,][]{espaillat2010}. \cite{pinilla2016} showed that this NIR excess emission can be explained at any time of evolution when the embedded planet is not more massive than $\sim1-5\,M_{\rm{Jup}}$ around a solar-type star (with a disk viscosity of $10^{-3}$). The gap formed by such a planet allows dust particles to pass throughout the gap and refill the inner disk with dust particles. In CIDA\,1, there is a very bright inner emission in both Band 7 and Band 4, which can help to constrain the mass and position of the potential planet creating the large gap. 

Under the assumption that it is one single planet creating the large gap, \cite{pinilla2018} analytically calculated the minimum mass of the planet needed to open a gap in the gas surface density under the conditions of CIDA\,1 and create a ring at 20\,au. Our current observations show that the minimum of this gap is around 8\,au. Assuming this location for the position of the potential planet, the minimum mass of the planet to open a gap in a disk with $\alpha_{\rm{visc}}=10^{-3}$ is $\sim 1.2\,M_{\rm{Saturn}}$ (or a planet-to-star mass ratio of $3.2\times10^{-3}$). When assuming such a planet at 8\,au, the pressure maximum is located at $\sim13$\,au. Hence either the planet must be more massive to create a wider gap and move the pressure maximum further out, or the planet should be located further out. In the latter case, the minimum of the gap seen in millimeter emission would not correspond to the location of the planet. This is a typical assumption in most hydrodynamical simulations that aim to explain some observational properties of protoplanetary disks, where it is assumed that the minimum of the dust emission co-locates with the minimum of the gas surface density, i.e., the local minimum of a gap opened by a planet.

When forcing the planet to be located at 8\,au, a planet-to-star mass ratio of $2.3\times10^{-2}$ is needed to have the pressure maximum at $\sim$20\,au, where the peak of the emission is located. This corresponds to a planet mass of $\sim2.4$\,$M_{\rm{Jup}}$ for CIDA\,1. Such a planet is unlikely to form by core accretion around such a low-mass star and in a disk of low mass \citep{liu2020}. As an experiment, we performed dust evolution simulations assuming a gap created by a $\sim2.4$\,$M_{\rm{Jup}}$ located at 8\,au. The shape of the gap is obtained analytically using the minimum mass to open a gap criterion by \cite{crida2006} and following \cite{pinilla2015}. For this simulation, we assumed an initial dust-to-gas ratio of 1/100 where all the grains are initially micron-sized particles. We used the dust evolution package DustPy (Stammer \& Birnstiel in prep). The simulations include the growth, fragmentation, and erosion of particles. The fragmentation velocity of the particles is 10\,m\,s$^{-1}$ in the entire disk. We assumed  $\alpha_{\rm{visc}}=10^{-3}$ and a disk mass of 5.5\,$M_{\rm{Jup}}$ around a stellar object of 0.1\,$M_\odot$ mass (i.e., $M_{\rm{disk}}/M_{\rm{CIDA1}}\sim 0.05$). For more details of the dust evolution models see \cite{birnstiel2010}. 

Figure~\ref{fig:dust_evo2} shows the dust density distribution as a function of grain size and distance from  the star for the case of a $\sim2.4$\,$M_{\rm{Jup}}$ located at 8\,au and three different times of evolution (0.1, 0.5, and 1\,Myr). The white line corresponds to the particle size for which the Stokes number (St) is equal to unity ($\rm{St}=\frac{a\rho_s}{\Sigma_g}\frac{\pi}{2}$), and therefore it also represents the shape of the gas surface density. These results show how the inner disk is depleted, because in this case the gap formed by such a massive planet is so deep that it blocks all  dust particles, and thus there is no replenishment of dust particles from the outer part to the inner part of the disk. As a result, an empty cavity is created, and already at 0.1\,Myr of evolution almost all the grains that were initially within the inner edge of the gap have grown and drifted towards the star.

This experiment demonstrates that such a massive planet is unlikely to exist in this disk, not only because its formation challenges current models of planet formation, but also because the new observational evidence of the inner disk at millimeter wavelengths excludes this possibility. 

As a potential solution, we performed an experiment moving the location of the planet to 12\,au, which allows us to have a lower mass planet to form a pressure maximum at 20\,au.  This situation relaxes the constraint of having the planet sitting at the minimum of the gap at 8\,au. The planet-to-star mass ratio in this case is $\sim5\times10^{-3}$, which corresponds to $\sim0.5\,M_{\rm{Jup}}$ around a 0.1\,$M_\odot$ stellar object.  

\begin{figure*}
    \centering
    \includegraphics[width=18.0cm]{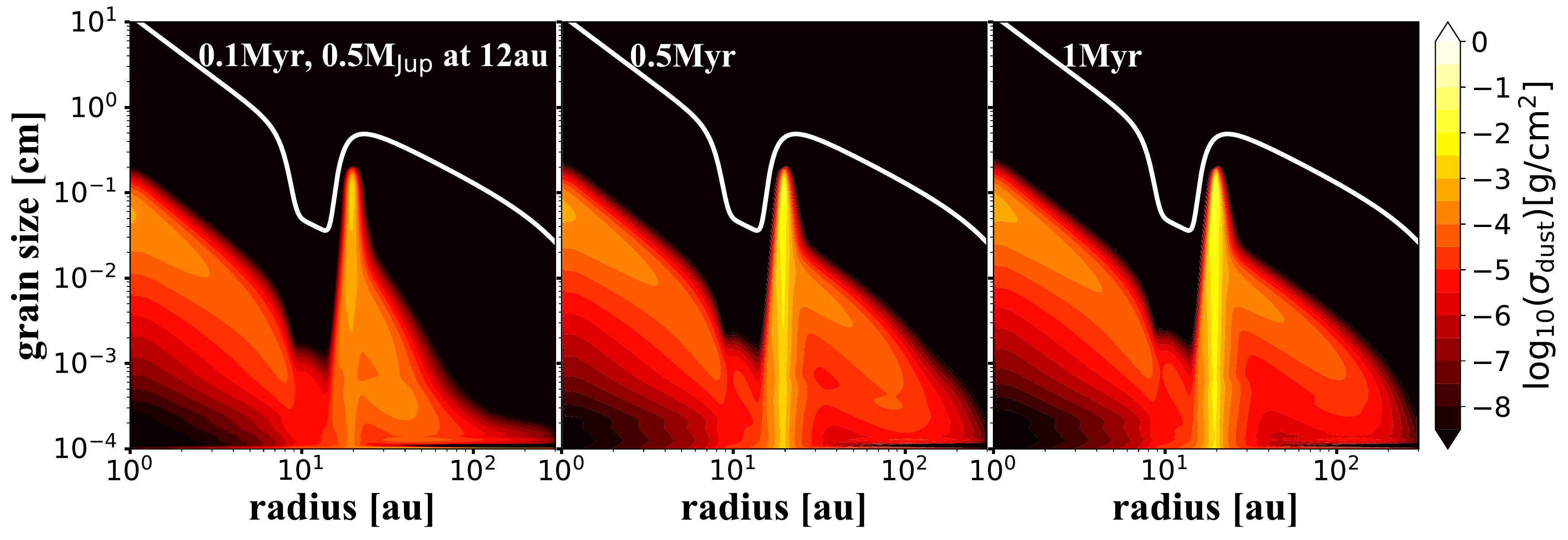}
    \caption{Particle density distribution after 0.1\,Myr, 0.5\,Myr, and 1\,Myr of evolution as a function of distance from the star and grain size. The white solid line represents a Stokes number of unity. This model assumes the influence of a $0.5$\,$M_{\rm{Jup}}$ planet located at 12\,au in the disk.}
    \label{fig:dust_evo}
\end{figure*}

\begin{figure*}
    \centering
    \includegraphics[width=18.0cm]{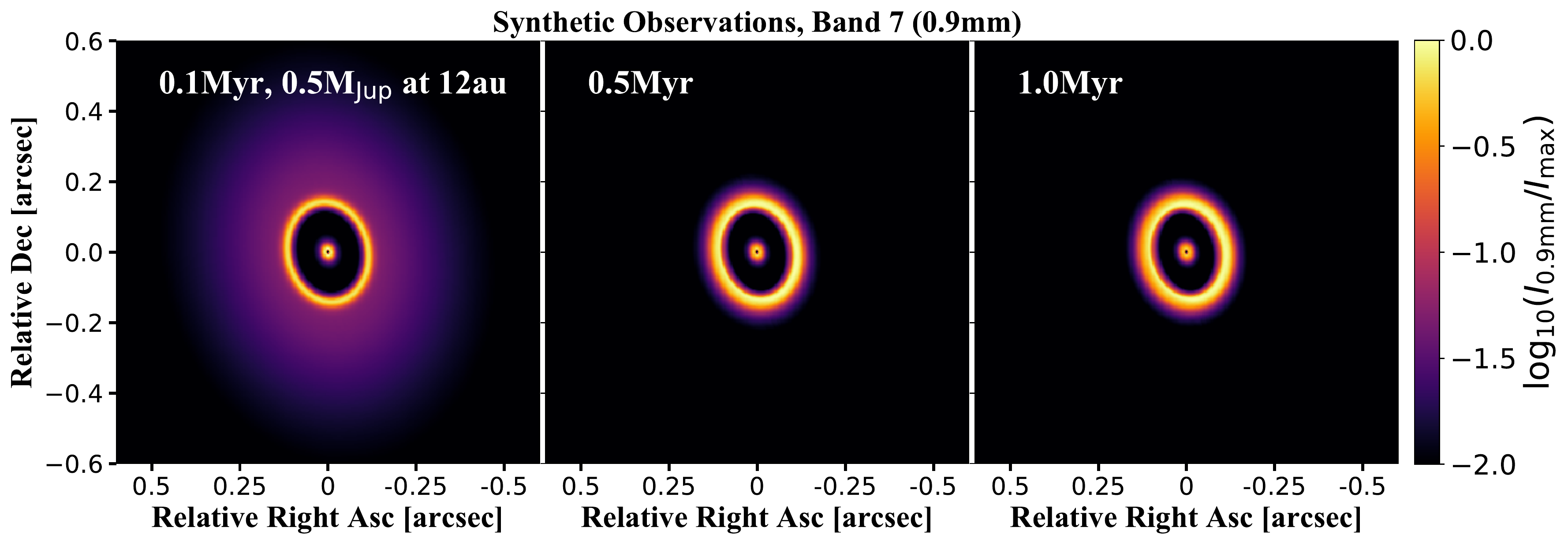}\\
    \includegraphics[width=18.0cm]{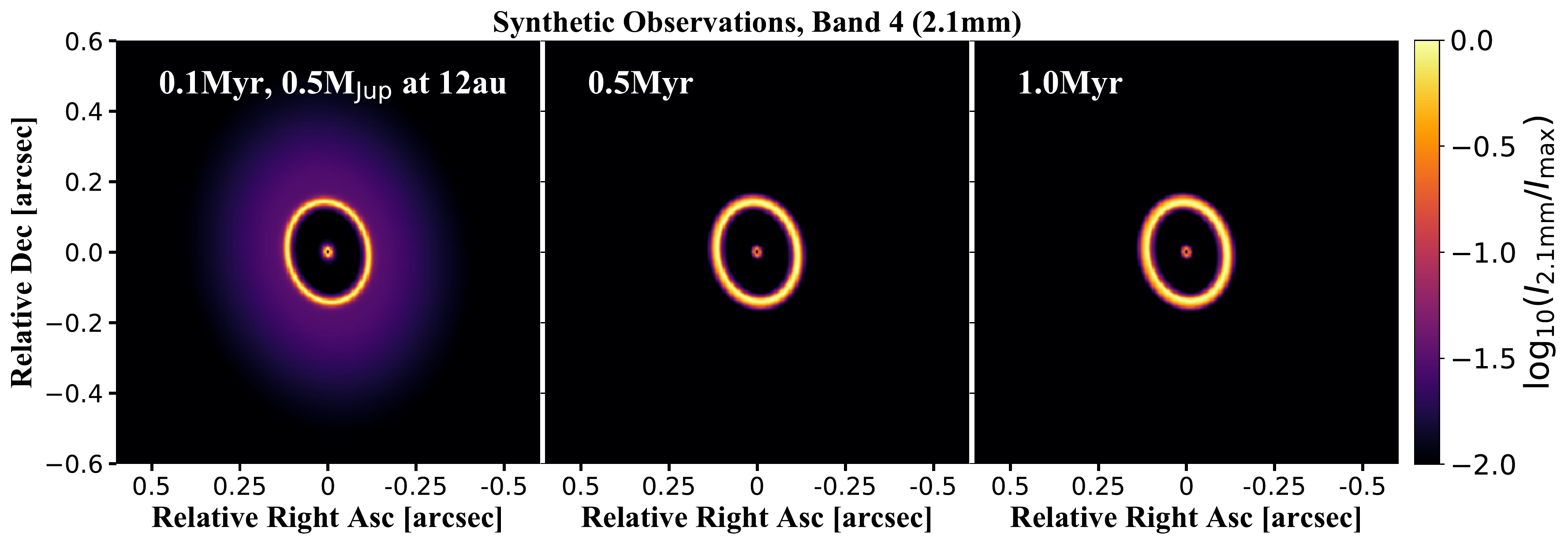}
    \caption{Synthetic observations at 0.9\,mm (top) and 2.1\,mm (bottom) from dust evolution models and radiative transfer calculations after 0.1\,Myr, 0.5\,Myr, and 1.0\,Myr of evolution for the case of one 0.5\,$M_{\rm{Jup}}$ planet at 12\,au creating a pressure bump at $\sim20$\,au.}
    \label{fig:synthetic_images}
\end{figure*}

Figure~\ref{fig:dust_evo} shows the dust density distribution for the case of a 0.5\,$M_{\rm{Jup}}$ at 12\,au at 0.1, 0.5, and 1.0\,Myr of evolution. The dust density distribution of the dust evolution simulations are then combined with radiative transfer calculations using the Monte Carlo code RADMC-3D\footnote{Code available at: \url{http://www.ita.uni-heidelberg.de/~dullemond/software/radmc-3d/}} as explained in previous works \citep[e.g.,][]{pohl2016}. Figure~\ref{fig:synthetic_images} shows the synthetic images at each time of evolution and for both wavelengths  (without convolution to clearly see the differences). The dust density distribution shows that in this case  the gap formed by this planet allows dust particles from the outer disk to move throughout the gap and replenish the inner disk. This inner dust-rich disk survives even after 1 Myr. With time, the dust density increases inside the ring because particles that are growing in the outer parts of the disk move towards the pressure maximum, which leads to higher intensity differences between the ring and the inner disk with time.

\begin{figure*}
   \tabcolsep=0.05cm 
   \begin{tabular}{cc}
   \centering
    \includegraphics[width=\columnwidth]{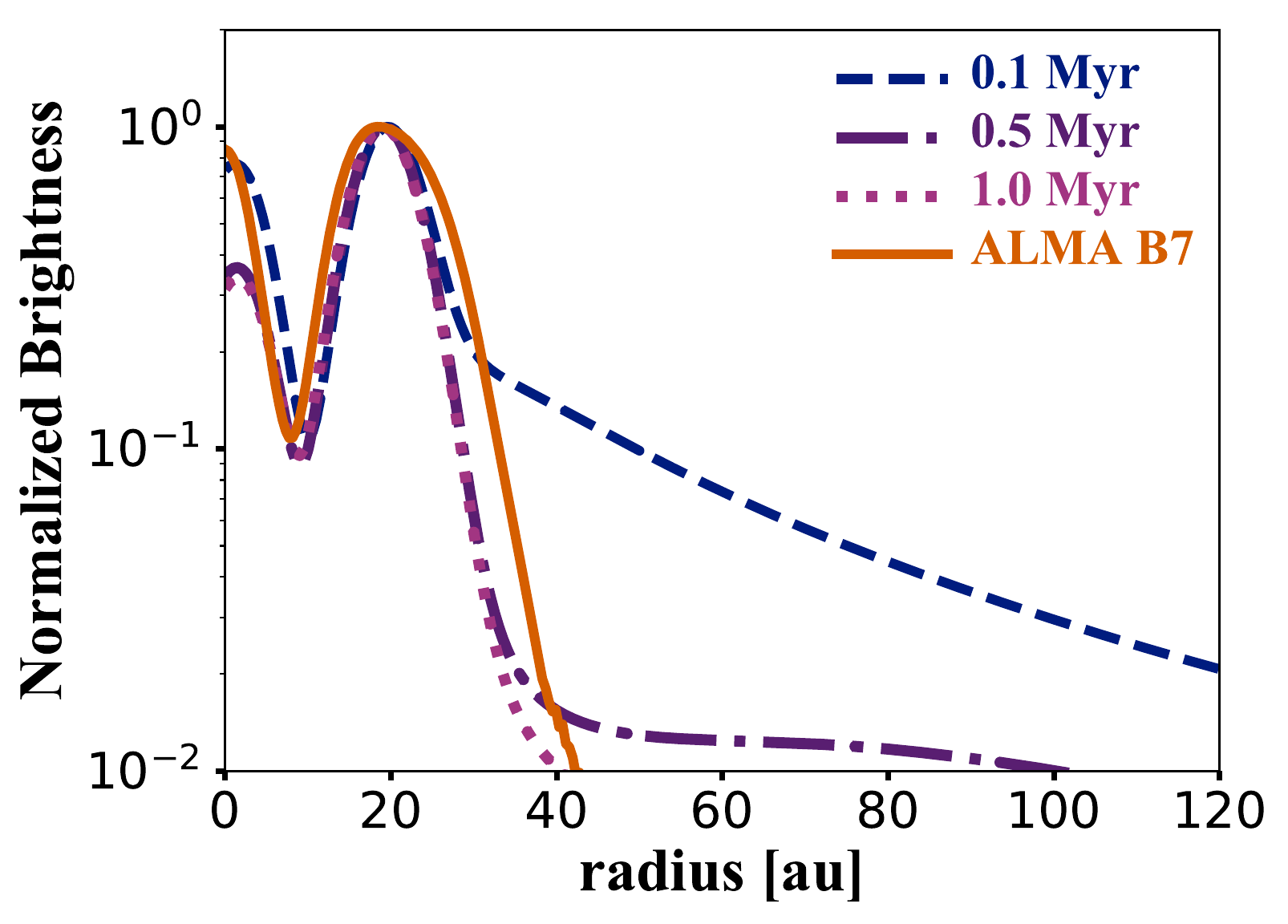}&
    \includegraphics[width=\columnwidth]{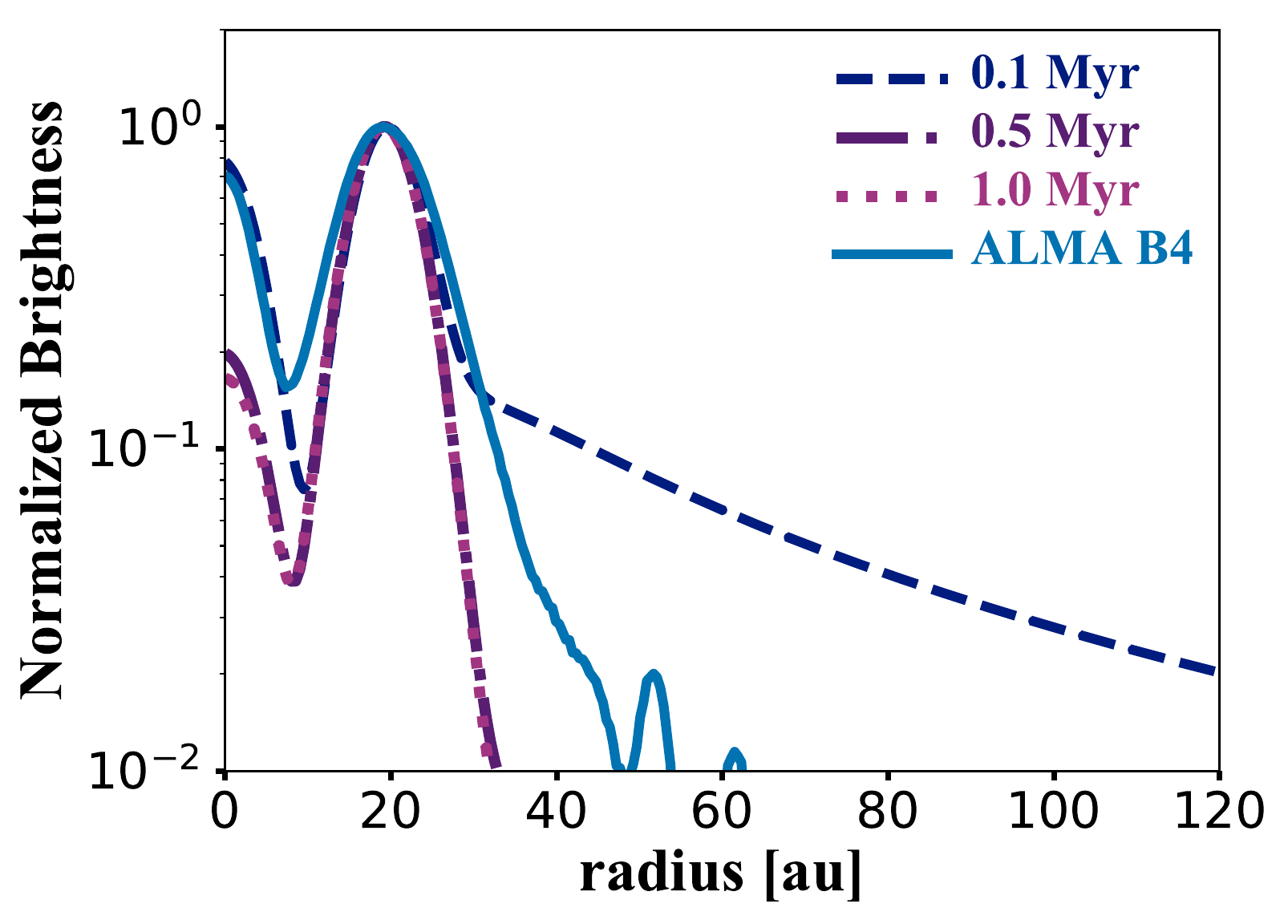}
    \end{tabular}
    \caption{Comparison  of the deprojected radial profile from Fig.~\ref{fig:radial_profile} and the models from Fig.~\ref{fig:dust_evo} (0.5\,$M_{\rm{Jup}}$ planet at 12\,au) after deprojecting and convolving with a beam similar to the one from observations.}
    \label{fig:rad_profile_LMP}
\end{figure*}

The synthetic images shown in Fig.~\ref{fig:synthetic_images} are normalized to the peak of the emission and they are in logarithmic scale. As seen in these images and in Fig.~\ref{fig:dust_evo2}, there is always an inner emission. The outer tail is only visible in the images at 0.1\,Myr of evolution and at 0.5\,Myr and 1.0\,Myr only the inner emission and the outer ring peaking at 20\,au remain. Figure~\ref{fig:rad_profile_LMP} shows the comparison of the deprojected radial profile from Fig.~\ref{fig:radial_profile} and the models  assuming a 0.5\,$M_{\rm{Jup}}$ planet at 12\,au after deprojecting and convolving with a beam similar to the one from observations. This comparison of the radial profiles illustrates how the models at 0.1\,Myr nicely reproduce the contrast between the inner disk and the peak of the ring, and the width and depth of the gap in Band 7. The minimum of the gap ($\sim8$\,au) is also in good agreement, even though  the location of the planet is set to 12\,au. In Band 4, the depth of the gap is higher than observed, although our visibility analysis suggests that the gap is shallower, in better agreement with these models. The main discrepancy between the radial profile at 0.1\,Myr and the observations is the bright and large tail in the outer disk that is not seen in the radial profile of the observations. At later times of evolution, while the outer tail vanishes, the inner disk also becomes less bright with respect to the peak. The discrepancy in the inner disk brightness between the observations and models is more noticeable for Band\,4. Therefore, these models cannot explain the very bright inner disk at 1\,My  as observed in the absence of a mechanism helping to slow down the radial drift of the particles in the inner disk. One alternative is that this planet gradually grows and reaches its final mass in the last 0.1\,Myr, such that the inner disk may still be as bright as observed.

To summarize these results of dust evolution models and comparison with observations, a very massive planet (as massive or more massive than Jupiter) is excluded because the gap would filter grains of all sizes and the inner disk would vanish very quickly. A $0.5\,M_{\rm{Jup}}$ allows replenishment of dust from the outer disk, which leads to good agreement between the contrast of the inner disk and the ring at early times of evolution (0.1\,Myr). At longer times, it is challenging to keep this bright inner disk. 

\section{Discussion} \label{sect:discussion}

The high accretion rate and the large cavity of the disk of CIDA\,1 exclude models of photo-evaporation \citep{ercolano2017} as a potential explanation for the origin of the cavity. Moreover, the new detection of the inner disk with ALMA provides further support to exclusion of photoevaporation as the mechanism to open this cavity. A possible alternative is the dust trapping that is expected at the outer edge of a dead zone \citep{flock2015}. However, as discussed in \cite{pinilla2018} most of the nonideal  MHD models have focused on T-Tauri or Herbig stars and little is known about how large  a dead zone 
can be around a VLMS such as CIDA\,1. In the remainder of the discussion, we focus on the possibility that the cavity is opened by a giant planet as in the previous section.

\subsection{Inner disk}
One of the main results from our observations is the detection of an unresolved inner disk in CIDA\,1. Recent observations with ALMA revealed unresolved inner disk emission in several transition disks \citep[e.g., T\,Cha, SR\,24S, HD\,100546, GG\,Tau, HD\,100453,][respectively]{hendler2018, pinilla2019, perez2020, phuong2020, rosotti2020}, see also \cite{francis2020}.

As discussed in the previous section, the existence of these inner disks can provide upper limits for the potential embedded planets in cavities, because if the planet-to-star mass ratio is very high ($\gtrsim 5\times10^{-3}-1\times10^{-2}$), the inner disk vanishes before 1\,Myr. In the case of CIDA\,1, the inner disk is very bright. Our visibility fits suggest that the ring peak is only $\sim$30\% of the peak of the inner disk emission (in Band 7) and in addition our spectral index analysis suggests that the inner disk and outer ring share similar values, which implies that both regions are rich in millimeter-sized particles assuming optically thin emission. This assumption of optically thin emission is probably inaccurate in the very hot inner parts of the disk (we obtained 160\,K at 1\,au from our radiative transfer calculations), in which case the spectral index does not provide any information about the grain size. 

In our dust evolution models we have difficulty  simultaneously explaining the inner disk (in particular its brightness) and the formation of the ring by one single massive planet. A less massive planet can help to have less dust filtration at the outer edge of the gap and hence a brighter inner disk with a higher replenishment of dust from the outer disk; but in such a case the pressure bump and the location of the ring would not be as far from the star as what is observed (at 20\,au). 

Our results also suggest that there may be a physical mechanism that helps to keep the inner disk brighter on million-year timescales. In our models, the grains that pass the gap and reach the inner disk are small dust particles (micron-sized particles), which then grow again and quickly drift towards the star. Hence, it is necessary to stop or slow down the radial drift of these dust particles in the inner disk to keep it bright.  To slow down the radial drift of particles in the inner part of disks, \cite{pinilla2016} considered the effect of the snow line in their dust evolution models. At the  location of the snow line, small particles can be recreated due to the lower fragmentation velocities that are expected from dry grain particles compared to icy dust particles \citep[see however recent results from laboratory experiments of dust collisions;][]{gundlach2018, musiolik2019, steinpilz2019}. The inclusion of variation of the fragmentation velocity at the snow line \citep[see also][]{garate2020} can explain the NIR excess from the SEDs seen in a subset of transition disks (pre-transition disks). However, in the simulations from \cite{pinilla2016}, this snow-line effect does not produce a significant inner disk at submillimeter emission (see their figure A.1).

Another possibility is to trap the particles at the inner edge of a dead zone \citep[e.g.,][]{dzyurkevich2010, hu2016, ueda2019}. Very close to the star where the temperatures are similar to the dust sublimation temperatures, the magnetorotational instability (MRI) is expected to operate due to thermal ionization. The inner edge of a dead zone has been proposed to be an excellent location for trapping of pebbles and for the formation of close-in planets. For the luminosity of CIDA\,1, it is expected that the dead-zone inner edge is at about 0.01\,au, similar to the case of Trappist\,1 and the location of the closest-in planet in that system \citep[see Fig.~8 from][]{flock2019}. However, if the grains are too small near the inner edge of the dead zone, trapping is not expected \citep{jankovic2019}, which can be the case inside the snow line.

Observations of  exoplanets suggest that cold Jupiters may be more common when they are accompanied by a closer-in super-Earth \citep[e.g.,][]{zhuwu2018, bryan2019}. If the large cavities observed in transition disks are the product of cold Jupiters embedded in the disk, it is possible that the inner disks remain due to the presence of an inner pressure bump created by a super-Earth, but detailed modeling is required to test this idea. 

Alternatively, in the context of planets creating the cavity, it is possible that there are multiple planets whose gaps overlap and create a wider but shallower gap \citep{duffell2015}. In this case, it may be possible to still have a bright inner disk over longer timescales of evolution, an idea that needs confirmation with hydrodynamical simulations of multiple planets, dust evolution models, and radiative transfer calculations.

\subsection{Multiple rings or tail?} 

Our results from the visibility fits indicate that there is outer emission outside the main ring of the CIDA\,1 disk. Our \texttt{galario} fits are in favor of an extra Gaussian on top of the main ring to reproduce that outer emission, which is only required in Band\,7 and is negligible in the best model of the fit of the Band\,4 observations. These results suggest that there is an outer shoulder in Band\,7 or that the ring is actually a composition of two (or more) unresolved rings. 

As discussed in the previous section,  we expect a tail of emission in the outer disk from dust evolution models when a planet is present, which is significant compared to the ring at early times of evolution. The tail in the dust density distribution does not disappear at longer times, but because the ring becomes brighter due to the drift of millimeter-sized particles in the outer disk, the emission of the tail becomes negligible with time. This emission in the models also has a different nature from what our analysis of the visibilities suggests. In the dust evolution models, it looks more like extended emission towards the outer disk (see Fig.~\ref{fig:rad_profile_LMP} and  Fig.~\ref{fig:synthetic_images}), while the best-fit models from visibility fitting using Band 7 data suggest that it looks more like an unresolved ring. 

Recent observations with ALMA of several transition disks showed that what was known to be a wide ring around a cavity is actually a composition of several narrow rings \citep[e.g.,][]{perez2020, facchini2020}. A couple of ideas have been proposed to explain these narrow rings. \cite{perez2020} suggested that a single migrating low-mass planet ($10\,M_\oplus$) in between these rings can explain the multiple rings and their separation, but this planet cannot explain the formation of the cavity itself. \cite{facchini2020} demonstrated that the appearance of one ring or multiple rings  depends on the assumptions of the thermodynamics in hydrodynamical simulations where a planet is assumed to be embedded inside the cavity. Alternatively, low values of the disk viscosity, both in hydrodynamical simulations and dust evolution models, can change the appearance of the disk that is hosting a giant planet, spanning appearances that suggest a compact disk all the way to those suggesting multiple rings and gaps \citep[e.g.,][]{ovelar2016, bae2018}. 

In the case of CIDA\,1, it remains unclear as to whether or not this outer ring is a composition of multiple structures, and deeper observations at higher angular resolution are required to solve this problem and to explore potential scenarios. 

\subsection{Spectral index}
Our observations suggest low average values of the spectral index of $\sim2$. According to our calculations, the  emission along the ring is mostly optically thin, which suggests that this low spectral index is due to a large number of millimeter-sized particles in the ring (the optical depth at the ring location from our radiative transfer models is much lower than unity). However, from the theoretical point of view of dust evolution models, it is challenging to reproduce these low values of the spectral index in disks around low-mass stars. This results from a combination of different factors. For example, radial drift velocities are higher around VLMSs than around T-tauri stars. This influences the spectral index in two different ways: (i) The millimeter-sized particles are lost towards the star on shorter timescales and (ii) drift mainly sets the maximum grain size of the particles \citep{pinilla2013}. This last element together with these disks being lower in mass leads to a maximum grain size that is at most 1\,cm in the inner disk (and usually decreases outwards), even when pressure bumps are present  (see Fig~\ref{fig:dust_evo}).  This problem of having a low growth barrier, in particular in disks around VLMSs, is worse when taking into account the recent results from laboratory experiments that suggest fragmentation velocities of icy particles of 1\,m\,s$^{-1}$ or even less \citep[e.g.,][]{musiolik2019}.

\begin{figure}
    \centering
    \includegraphics[width=9.0cm]{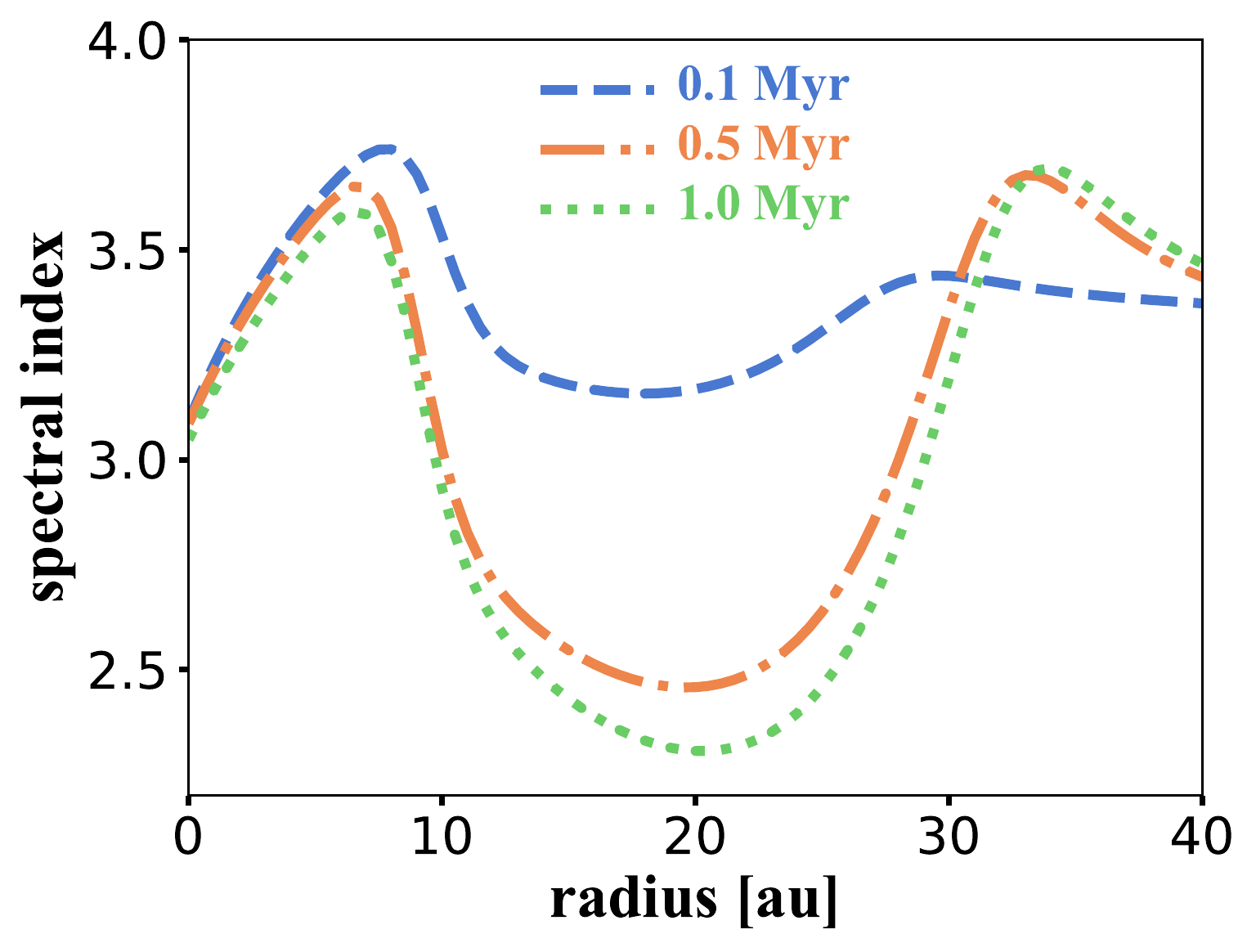}
    \caption{Spectral index profile obtained from the intensity
profiles from models that assume one single planet of $0.5\,M_{\rm{Jup}}$ at 12\,au.}
    \label{fig:spectral_index_model}
\end{figure}

 Dust evolution simulations of disks around VLMSs, with one planet creating a bump and trapping the grains in a single ring, fail in reproducing very low spectral index values as observed in CIDA\,1. Figure~\ref{fig:spectral_index_model} shows the spectral index obtained from the intensity profiles  from models that assume one single planet of $0.5\,M_{\rm{Jup}}$ at 12\,au (Sect.~\ref{sect:models_obs}). It is important to note that these models already include the full treatment of the dust scattering, which is known to decrease the spectral index values \citep{liu2019}
 
 In this figure, it is possible to see how the spectral index is expected to decrease inside the ring to values as low as 2.5 and to values of $\sim$2.6 when integrating in the entire disk. At 1\,Myr of evolution, 20\% of the dust has been lost due to drift in the dust evolution model that considers a $0.5\,M_{\rm{Jup}}$ at 12\,au, but a very small fraction of the remaining mass is in millimeter-sized particles. The dust mass in particles larger than 0.1\,mm is only 0.3\,$M_\oplus$, while the dust mass in the smaller grains is  13.7$M_\oplus$, which means that only 2\% of the mass is in large grains ($\geq 0.1$\,mm). From simple radiative transfer calculations, a way to mitigate this problem is to change the fraction of small versus large grains, which already provides values of the spectral index as observed (i.e., $\sim2$). To have this switch in our dust evolution models, we would need multiple and strong pressure bumps in the entire disk that allow to accumulate a lot of millimeter-sized particles in the whole disk \citep[as in][]{pinilla2013}. However, our high-angular-resolution observations only suggest the presence of one single particle trap and therefore it remains unclear as to how we can explain the low spectral index of CIDA\,1. This will require further investigation with dust evolution models to search for the set of parameters that allows us to have the large majority of the dust in large particles, including fragmentation velocities and gas and dust diffusion parameters.

\section{Conclusions} \label{sect:conclusions}
In this paper we present new ALMA and X-Shooter observations of the disk around CIDA\,1 (0.1-0.2\,$M_\odot$), which is one of the very few known disks that hosts a large cavity (20\,au in size) around a very low mass star. Our ALMA observations include  Band\,7 (0.9\,mm) and Band\,4 (2.1\,mm) both with a resolution of $\sim 0.05''\times 0.034''$. Our main findings can be summarized as follows.

\begin{itemize}

\item The X-Shooter observations confirm the  accretion rate of CIDA\,1 of $\dot{M}_{\rm acc}$ = 1.4 $~\times~10^{-8} M_\odot$/yr for its stellar mass. This high value of $\dot{M}_{\rm acc}$ and the large cavity of 20\,au exclude  photo-evaporation as a potential origin of the observed structures.  

\item The new ALMA observations reveal an unresolved inner disk in the continuum emission at the two wavelengths. This inner disk is very bright when compared with the emission of the main ring, which peaks at 20\,au. The discovery of this inner disk helps to constrain models for the formation of the observed cavity. 

\item Our experiments with dust evolution models that include a massive planet in the disk   reveal an inner disk that is fainter than observations suggest for this system, demonstrating that a massive planet ($\gtrsim0.5\,M_{\rm{Jup}}$) is unlikely to exist in this disk. Our models of dust evolution with a $0.5\,M_{\rm{Jup}}$ planet at 12\,au can explain the contrast of the inner disk and ring only at 0.1\,Myr of evolution. After 1 Myr the inner disk becomes fainter because a nonphysical mechanism stops the radial drift of the particles in the inner disk. A potential alternative is that the planet around CIDA\,1 has gradually grown and reached its final mass in the last~0.1 Myr.

\item There are additional possibilities to explain a long-lived inner disk as observed with ALMA for CIDA\,1. These include a traffic jam expected at the snow line due to variation of fragmentation velocities of the particles, the inner edge of dead zones, and/or multiple planets inside the cavity; e.g., a super-Earth in the very inner part of the disk.  The effects of the inclusion of these scenarios in planet--disk interaction and dust evolution models remain to be explored.

\item Besides the inner disk and the ring-like structures, there is no further substructure found in CIDA\,1 at the current resolution and sensitivity. There is a hint of a shoulder beyond the ring-like structure from our fit of the visibilities in Band\,7. It is possible that the ring is composed of two or more narrow rings as observed in other transition disks, and higher angular resolution observations are needed to verify this possibility. 

\item There is weak evidence of a decrease in the spectral index inside the ring, both from observations and from our fits of the visibilities. The spatially integrated spectral index is very low ($\sim
2$), which is very challenging to explain in dust evolution models. Better agreement between models and observations is obtained if  most of the dust inside the ring of the models remains as millimeter-sized particles, which calls into question our understanding of the different physical processes, such as growth, fragmentation, and diffusion inside pressure bumps. 

\end{itemize}

\section*{Acknowledgements}
P.P. and N.T.K. acknowledge support provided by the Alexander von Humboldt Foundation in the framework of the Sofja Kovalevskaja Award endowed by the Federal Ministry of Education and Research. This work was partly supported by the Deutsche Forschungsgemeinschaft (DFG, German Research Foundation) – Ref no. FOR 2634/1 TE 1024/1-1.
M.T. has been supported by the UK Science and Technology research Council (STFC) via the consolidated grant ST/S000623/1, and by the European Union’s Horizon 2020 research and innovation programme under the Marie Sklodowska-Curie grant agreement No. 823823 (RISE DUSTBUSTERS project). S.M.S. acknowledges funding from the European Research Council (ERC) under the European Unions Horizon 2020 research and innovation programme under grant agreement No. 714769, the Deutsche Forschungsgemeinschaft (DFG, German Research Foundation) under Germany’s Excellence Strategy – EXC-2094 – 390783311, and the support from the DFG Research Unit “Transition Disks” (FOR 2634/1). This paper makes use of the following ALMA data: ADS/JAO.ALMA\#2018.1.00536.S, ADS/JAO.ALMA\#2015.1.00934.S, and ADS/JAO.ALMA\#2016.1.01511.S.  ALMA is a partnership of ESO (representing its member states), NSF (USA) and NINS (Japan), together with NRC (Canada), MOST and ASIAA (Taiwan), and KASI (Republic of Korea), in cooperation with the Republic of Chile. The Joint ALMA Observatory is operated by ESO, AUI/NRAO and NAOJ.
  

\bibliographystyle{aa} 
\bibliography{CIDA1main.bbl}

\end{document}